\documentclass[12pt,preprint]{aastex}
\usepackage{epsfig}
\usepackage{amsmath}

\begin{document}
%\title{The Distribution of Cool Atomic Gas in the Inner Milky Way:
%H{\tt I} Absorption in the SGPS Test Region}

\title{Fitting Together the HI Absorption and Emission in the SGPS}

\author{John M. Dickey}
\affil{Department of Astronomy, University of Minnesota, 116 Church St. SE,
Minneapolis, MN  55455; john@astro.umn.edu}
\author{N. M. McClure-Griffiths}
\affil{Australia Telescope National Facility, PO Box 76, Epping, NSW 1710;
Australia, Naomi.McClure-Griffiths@atnf.csiro.au}
\author{B. M. Gaensler}
\affil{Department of Astronomy, Harvard University,
60 Garden St. MS-6,Cambridge MA 02138; bgaensler@cfa.harvard.edu}
\author{A. J. Green}
\affil{School of Physics, University of Sydney,
NSW 2006, Australia; agreen@physics.usyd.edu.au}

\begin{abstract}

In this paper we study 21-cm absorption spectra
and the corresponding emission spectra  
toward bright continuum sources in
the test region (326\arcdeg$< l <$ 333\arcdeg)
of the Southern Galactic Plane Survey.
This survey combines the high resolution of the
Australia Telescope Compact Array with the full
brightness temperature information of the Parkes single
dish telescope.
In particular, we focus on the abundance and temperature of the cool
atomic clouds in the inner galaxy.  The resulting
mean opacity of the H{\tt I}, $<\kappa>$, is measured as a function of
Galactic radius; it increases going in from the solar
circle, to a peak in the molecular ring of about 
four times its local value.  This suggests that the
cool phase is more abundant there, and colder, than
it is locally.  

The distribution of cool phase temperatures is derived
in three different ways.  The naive, ``spin temperature''
technique overestimates the cloud temperatures, as 
expected.  Using two alternative approaches we get 
good agreement on a histogram of the cloud temperatures,
$T_{cool}$, corrected for blending with warm phase gas.  The
median temperature is about 65 K, but there is a long
tail reaching down to temperatures below 20 K.  Clouds
with temperatures below 40 K are common, though not
as common as warmer clouds (40 to 100 K).

Using these results we discuss two related quantities,
the peak brightness temperature seen in emission 
surveys, and the incidence of clouds seen in H{\tt I}
self-absorption.  Both phenomena match what would
be expected based on our measurements of $<\kappa>$ 
and $T_{cool}$.

\end{abstract}


\section{Background}

Absorption by the 21 cm line of H{\tt I} was first detected 48 years
ago (Hagen, Lilley, and McClain, 1955), just four years after the 
line was detected in emission from interstellar hydrogen in
the Galactic plane.  As 
spectral line interferometers were developed in the 1960's
(Clark, Radhakrishnan, and Wilson 1963, Radhakrishnan et al. 1972a)
it became possible to measure absorption spectra toward many
more continuum sources at high and low latitudes.  The 
striking result from these pioneering studies was that the
H{\tt I} absorption spectra look very different from emission spectra
in the same directions (Clark 1965, Hughes, Thompson, and 
Colvin, 1971, Radhakrishnan et al. 1972b).  This can 
be understood as a result of the huge range of temperatures
in the interstellar hydrogen (Field 1958, Field, Goldsmith, and
Habing 1969).  Interstellar atomic hydrogen can exist in stable
thermodynamic equilibrium either at warm temperatures, $\sim$5000 to 
$\sim$10,000 K, or at cool temperatures, $\lesssim$50 
to $\sim$150 K; these two
phases are reflected in two regimes of excitation temperature
for the 21-cm line.  Generally this ``spin temperature'' 
is close to the kinetic
temperature of the gas, but see Liszt (2001) for a
detailed discussion of the exceptions to this rule.  There 
is good evidence that interstellar hydrogen is present at
temperatures intermediate between those of the classic 
warm neutral medium (WNM) and cool neutral medium (CNM)
(Heiles 2001), but generally the different morphology of
the emission and absorption spectra can be explained by
a mixture of gas along the line of sight at widely different
excitation temperatures (see Kulkarni and Heiles 1988 for a 
deeper discussion of the phases of the interstellar medium).
 
This paradigm for understanding the different structures of 21-cm
emission and absorption spectra as due entirely to variations
in the excitation temperature has been challenged occasionally
on the basis of the different physical volumes sampled by the
two kinds of spectra. The problem is not so much the necessity
of moving the telescope on and off the background continuum
source to obtain the absorption and emission spectra, respectively.
The real problem is the different solid angles contributing.
Emission at 21-cm is so widespread that
the emission spectrum always corresponds to the beam area of
the telescope, while the absorption spectrum comes only from
the gas in front of the continuum source.  The solid angles
of the two can differ by several orders of magnitude,
which raises the question of whether the differences between
emission and absorption could be due simply to small scale
structure in the interstellar medium (Faison and Goss 2001, 
Deshpande 2000).  To get
around this problem requires 
an emission-absorption study using the smallest possible
beam size for the emission and using background sources which
are relatively large, if possible the same size as the beam.
The new mosaic surveys of the Galactic plane, combining single
dish and interferometer data, allow this for the first time.
The Southern Galactic Plane Survey (SGPS, McClure-Griffiths et al.
2000) and the Canadian Galactic Plane Survey (CGPS, Taylor et al.
2002, Strasser et al. 2002) provide good quality emission-absorption
spectrum pairs using extended continuum sources at low latitudes,
with a beam small enough to mitigate the effects of variations
in the emission and absorption over small angles.

In this paper we concentrate on a small, test region of the SGPS 
(McClure-Griffiths et al. 2001) which
has been mapped at relatively high resolution (FWHM 90\arcsec).
Over the next year maps of the rest of the
$\sim$210 square degrees of the SGPS with similar resolution
will become available,
which will provide some 30 times the number of background
sources as those considered here.  The purpose of this paper
is to test methods of analysing the emission-absorption spectrum
pairs, and their interpretation.  The next section (2) discusses the
best way to obtain the spectra from the data.  We then consider 
what the absorption spectra alone tell us about the opacity of
the ISM in the inner galaxy (section 3).  Then comes
the tricky question of how best to combine
the information from the emission and absorption 
spectra to estimate the spin temperatures of the cool clouds (section 4).
We consider several fitting techniques that parallel approaches
used in past studies at intermediate latitudes.  Finally we 
discuss the implications of the spin temperature distribution 
for observable quantities like the peak brightness temperature
of the H{\tt I} emission and H{\tt I} self-absorption (section 5).
 
\section{How to Get the Emission and Absorption Spectra from the Data}

% paragraph below added Sept 2002 (ref point 10)...
The H{\tt I} 21-cm line is one of the only transitions in the entire
electromagnetic spectrum for which
emission and absorption are both relatively easy to detect from
the same region.  The two fundamental spectra are the brightness
temperature of the emission, $T_B(v)$, and the optical depth, $\tau (v)$.
Where there is no background continuum we observe $T_B$ directly,
and toward an extremely strong background source we observe $\tau$
with negligable contribution from the emission, but toward most 
continuum sources we see a mixture of emission and absorption. 
This section deals with the optimum method for obtaining $T_B$ and
$\tau$ from the survey data.

There are always more weak background sources than strong ones; 
to get as many absorption spectra as possible in a given area we need
to find the most effective way to extract the absorption spectra from
the data, since the fainter continuum sources give absorption spectra
of marginal quality.  For this experiment, as for most studies
of Galactic 21-cm emission and absorption, the limiting factor
which sets the noise level in the absorption spectrum
is not the radiometer noise but the precision with which we
can subtract the emission in the directions of the background
continuum sources, leaving the absorption only.  This is a
severe problem at low latitudes; the difficulty of doing
accurate emission subtraction has made single dish low latitude 
emission-absorption studies very difficult, even for Arecibo (Kuchar
and Bania, 1991).  

Interferometer studies offer such high resolution
that the emission can always be eliminated, at least toward the
most compact background sources, but 
lacking the zero-spacing (total power) information, interferometer
data alone cannot be used to find the emission spectrum.  So for
emission-absorption studies we need both single dish and
interferometer data.  The standard method for many years has 
been to measure the absorption spectrum with an interferometer
and compare with an emission spectrum from a single dish telescope
(Radhakrishnan et al. 1972c, Dickey et al. 1983, Mebold et al. 1982).
In the SGPS we have carefully combined single dish (Parkes) and
interferometer (ATCA\footnote{The Australia Telescope National Facility is 
funded by the Commonwealth of Australia for operation as
a National Facility managed by the Commonwealth Scientific and
Industrial Research Organisation}) data
to achieve complete {\it uv} coverage for
all angular scales larger than our synthesized beam size
(90\arcsec). Thus for absorption studies we have the best
of both worlds, with sufficient resolution to remove the
emission in the direction of the continuum sources to get
accurate absorption spectra at low latitudes, while still
preserving the total power information which gives us the
corresponding emission spectra.  
%sep02 
%We have not subtracted the
%continuum from the line data, as is often done early-on in
%aperture synthesis mapping, so we can convert easily to 
%optical depth by dividing by the off-line intensity in
%each spectrum.

\begin{figure}
%\epsscale{0.6}
%\plotone{fig1.ps}
\vspace{3in}
\epsfig{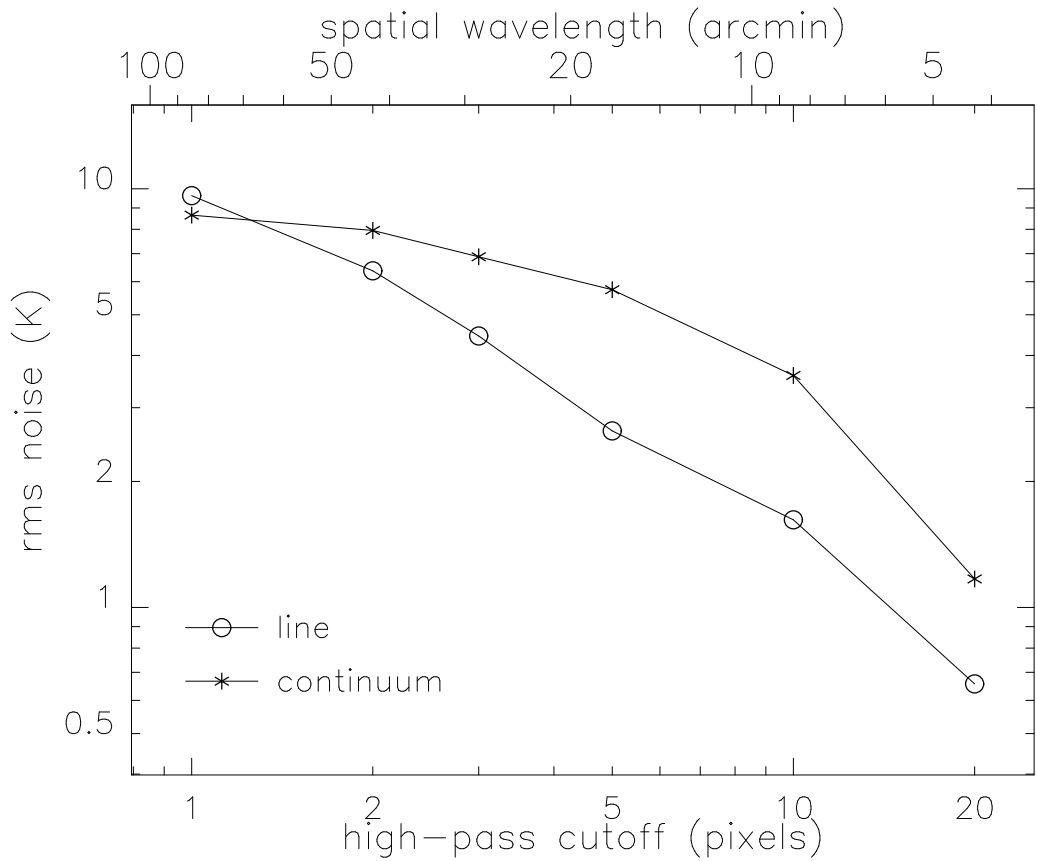}
\caption{The rms of the emission after high-pass spatial filtering.
The abscissa is the high-pass filter cutoff radius on the {\it uv}
plane, and the ordinate is the rms of the remaining emission.
The 21-cm line emission is shown by the circles, the continuum
is indicated by the stars.  The continuum is made up of a 
combination of diffuse emission and discrete sources, while
the line emission is mostly diffuse.  Thus filtering has a 
stronger effect on the line channels than on the continuum.
The pixel size in this case is 1.4 (deg)$^{-1}$; the corresponding
spatial wavelengths are shown on the top scale.}
\end{figure}

To obtain the most accurate estimate for the absorption spectrum
we still must remove the emission as completely as possible from
the spectrum toward the continuum source.  This can be done optimally
by considering the {\it uv} distribution of the continuum flux
vs. the emission line flux.  Generally the continuum is more
compact spatially than the line emission, which translates
to higher flux at longer {\it uv} spacings.  The 21-cm emission
%sep02 
follows a power-law distribution of flux vs. $\sqrt{u^2 + v^2}$ 
so that filtering in {\it uv} space allows us
to reduce the emission in a predictable way.  The filtering
is performed by Fourier transforming the cube to uv space,
then simply setting to zero all Fourier components within
a given radius of the origin, then transforming back.
This is illustrated on figure 1, which shows the rms
measured in the line channels vs. the uv radius of the filter
for a test region 1.42 \arcdeg square with little
continuum emission (region 1 of Dickey et al., 2001). 
The $uv$ pixel size in this case is 1.4 (degrees)$^{-1}$, meaning
that a filter radius of one uv pixel (which takes out just
the zero-spacing point) leaves spatial structure with sizes
smaller than $\sim$0.7$\arcdeg$, while a filter radius of ten
pixels leaves only emission in structures smaller than 
$\sim$4$\arcmin$.  Figure 1 shows that the rms level of the emission
remaining after filtering is about 10 K for filter width of
one pixel, decreasing to 1.7 K for the ten pixel filter.
Also shown on figure 1 is the continuum, using a region 
1.42 \arcdeg square centered on (l,b)=(331.43,0.56) that
includes both diffuse continuum and a few bright sources. 
The continuum is only
weakly affected by the filtering for filter widths less than about
five pixels, but progressively more strongly attenuated for
filter widths ten pixels or more.  
On the basis of figure 1 we choose a ten pixel filter
[passing spatial frequencies higher than (8.5 arc min)$^{-1}$
which preserves 
angular sizes smaller than about 4\arcmin], since
filtering more heavily begins to reduce the continuum as much as the
line emission.  Already with this filtering the total continuum 
flux of our background sources is less than 
it is on the unfiltered map, since they are extended
objects, thus we are beginning to lose signal to noise
on the absorption spectra.  We tabulate below these
filtered continuum peak values since they determine
the noise level in the absorption spectra, but these 
should not be used
to estimate the flux densities of the continuum sources.
%Figure 2 shows spectra taken toward a
%relatively faint continuum source (G331.27-0.19) after
%various levels of filtering.

In addition to removing the emission toward the continuum
sources to obtain the absorption spectra as described in
the last paragraph, we need to interpolate the emission 
from nearby beam areas in order to estimate the emission
spectrum which would be seen in the direction of the continuum
source if there were no absorption.  This we do by a spatial
interpolation on the unfiltered cube, which includes information
from all the short uv spacings.  The technique
is described by McClure-Griffiths et al. (2001); we construct
for each spectral channel a bi-linear function fitted to the
pixels around the background source for which the continuum 
brightness is below a threshold set at 20\% of its peak
value on-source.  We use the off-source pixels so defined 
both for the fitting and to estimate the error of the fitted
function; the latter is given by the rms of the data minus the fit
averaged over the set of off-source pixels.
That rms should be an overestimate of the
likely error in the interpolated spectrum in the direction
of the background source, since the area covered by the 
off-source pixels is much larger than that of the source
itself.  The error envelope defined this way is indicated
above and below the emission profiles on figures 2-8.
%sep02   all the rest of this paragraph added in response to ref point 2)
The absorption spectra are constructed by averaging spectra
toward the pixels for which the ratio of the continuum brightness 
to the continuum peak is above 80\%, weighting by the square of
this ratio.  
%sep02   all the rest of this paragraph added in response to ref point 2)
(The continuum map used to determine this weighting
has also been spatially filtered in the same way as the spectral line cube.)
Note that we do not need to subtract the interpolated
emission from the spectrum toward the continuum to obtain the
absorption spectrum, since the spatial filtering process has
already accomplished that step.  This is the difference between
the analysis performed here and that of McClure-Griffiths et al.
(2001).  Comparing the spectra in figure 8 of that paper with
the corresponding spectra in figures 2 and 3 below, we see that
the effects of emission fluctuations have been attenuated by a
factor of two to three by the spatial filtering step described
above.  The accuracy of both the interpolated emission spectrum
and the absorption spectrum is limited by this filtering and/or
interpolation step, as neither can be observed directly without
some pollution by the other.  Thus the matched-filter based on 
figure 1 is of central importance.
%sep02   end of addition, ref comment 2)

%\include{table1}

The brightest continuum sources in the SGPS test region are listed
on table 1, which gives the source name, rms noise
in optical depth ($\sigma_{\tau}$), integrated flux density in Jy,
peak brightness temperature in K, and recombination line velocity,
in columns 1 - 5.  Columns 6 - 10 give the values of various
spectral integrals defined below.
All of these continuum sources are H{\tt II} regions except for G328.42+0.22
which is a supernova remnant.
The H{\tt II} regions all have recombination lines
detected in the survey of Caswell and Haynes (1987);
the measured recombination line
center velocity is an indication of the kinematic
distance.  In a few cases the H{\tt I} absorption continues to
velocities beyond (i.e. more negative than) the recombination
line velocity.  This may be an indication that the H{\tt II} region
is at the far distance corresponding to its velocity 
(G328.82-0.08 is an example).  In some cases however the
H{\tt I} absorption spectrum shows a single deep component extending
5 to 10 km s$^{-1}$ beyond the recombination line velocity, and then
no absorption at higher velocities (G326.45+0.90 and G326.65+0.59
are examples).  These suggest local departures from the smooth
circular rotation described by the rotation curve, perhaps due
to the effect of the Sgr-Car arm at about -50 km s$^{-1}$.  The
recombination line velocities themselves may be offset from
the H{\tt II} region velocity due to asymetric expansion. 

In practice, peak continuum brightness temperatures of 
$T_c \ge 50$ K
or higher are needed to give high quality absorption spectra
($\sigma_{\tau} \le$0.03), but we include on table 1 all 13
sources with peak brightness above 20 K for the sake of the
line integrals discussed in the next section.  For the spin
temperature discussion in the subsequent section we will restrict
the analysis to the five sources with $T_c > 60$ K.  
Gaussian fitting is done on the nine with $T_c > 30K$. Spectra toward 
all the continuum sources listed on table 1 are shown on figures
2 - 8, with the interpolated emission above and the 
absorption below.  The intrinsic properties of most
of these sources are discussed
in detail by McClure-Griffiths et al. (2001).

\begin{figure}
%\epsscale{0.6}
%\plotone{fig2.ps}
\vspace{6.5in}
\epsfig{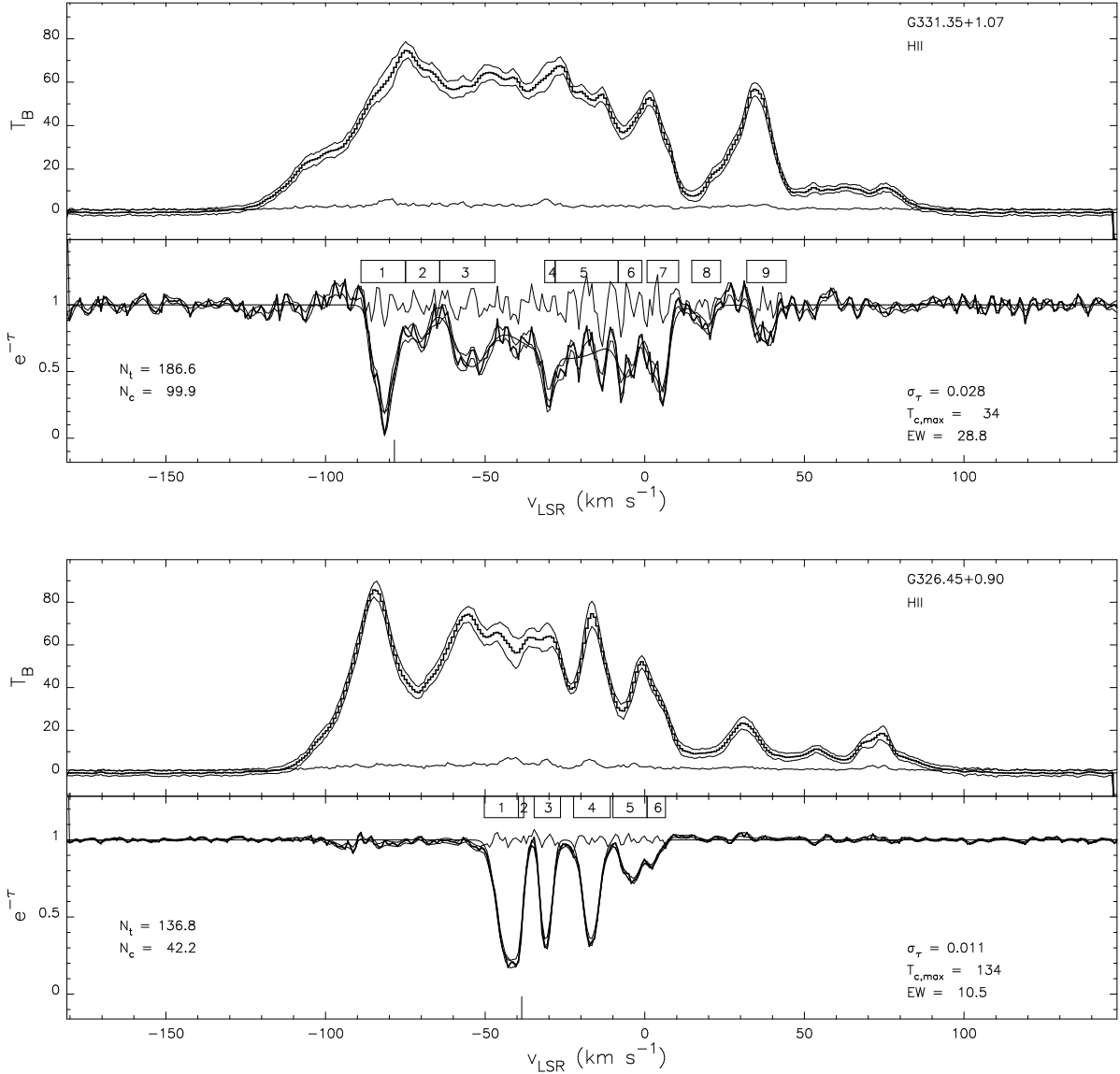}
\caption{Emission absorption spectrum pairs toward G331.35+1.07 (top)
and G326.45+0.90 (bottom).  For both, the emission spectrum is at the
top and the absorption spectrum is below.  Both spectra have error
envelopes, which are computed from the errors of the interpolation 
of the emission in the direction of the continuum source.
The numbered boxes in between
show the velocity ranges of the different blended lines.  The rms
noise in the absorption, coming primarily from the emission fluctuations,
is indicated as $\sigma_{\tau}$, the peak continuum brightness temperature
is indicated as $T_{c,max}$ (in K), and the equivalent width, EW is indicated
(in km s$^{-1}$).  Two values of the emission integral over negative 
velocities are shown, converted to units of 10$^{20}$ cm$^{-2}$, as 
discussed in the text.  Also shown are Gaussian fits to the absorption
(equation 9) and their residuals (data minus fit) for spectra toward
continuum sources with $T_{c,max} > 30$ K.}
\end{figure}

\begin{figure}
%\epsscale{0.6}
%\plotone{fig3.ps}
\vspace{7in}
\epsfig{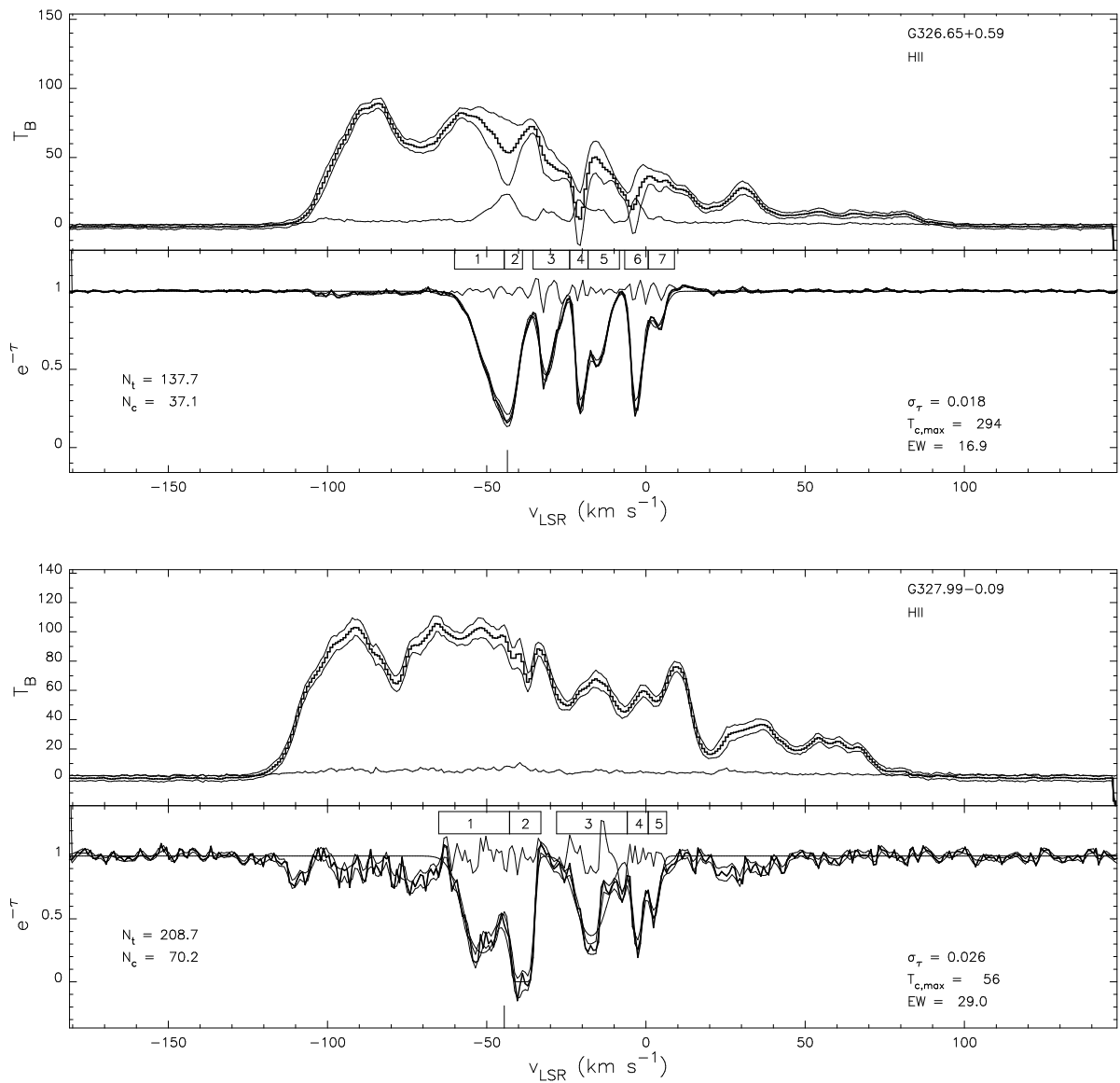}
\caption{Emission-absorption spectrum pairs for G326.65+0.59 and G327.99-0.09,
presented as in figure 2.  The absorption spectrum toward G326.65+0.59
was observed by Goss et al. (1972) in their pioneering study of Galactic
21-cm absorption.  Our spectra are consistent with theirs within the
uncertainties caused by emission fluctuations in their data.}
\end{figure}

\begin{figure}
%\epsscale{0.6}
%\plotone{fig4.ps}
\vspace{7in}
\epsfig{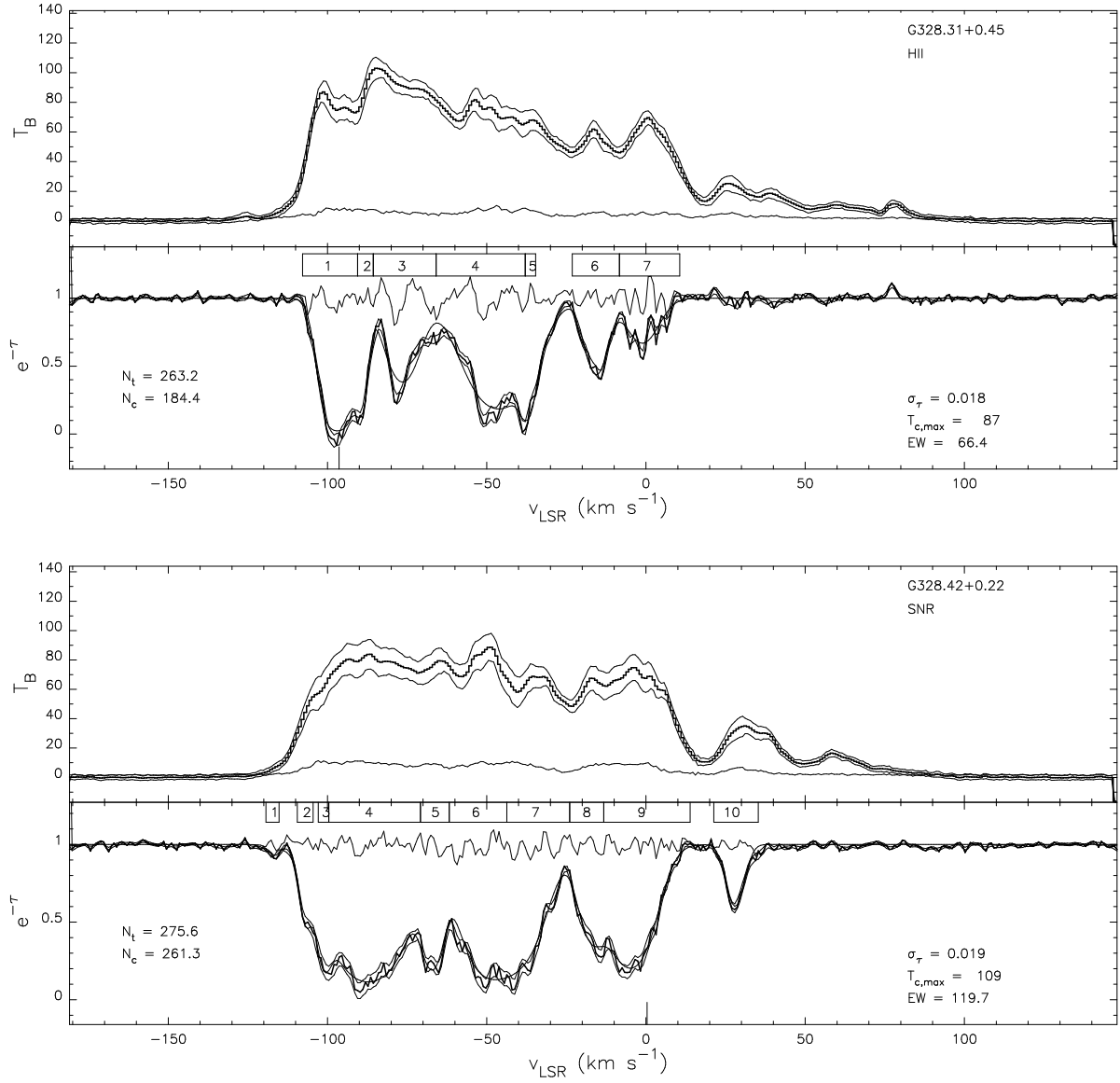}
\caption{Emission-absorption spectrum pairs for G328.31+0.45 and G328.42+0.22,
presented as in figure 2.}
\end{figure}

\begin{figure}
%\epsscale{0.6}
%\plotone{fig5.ps}
\vspace{7in}
\epsfig{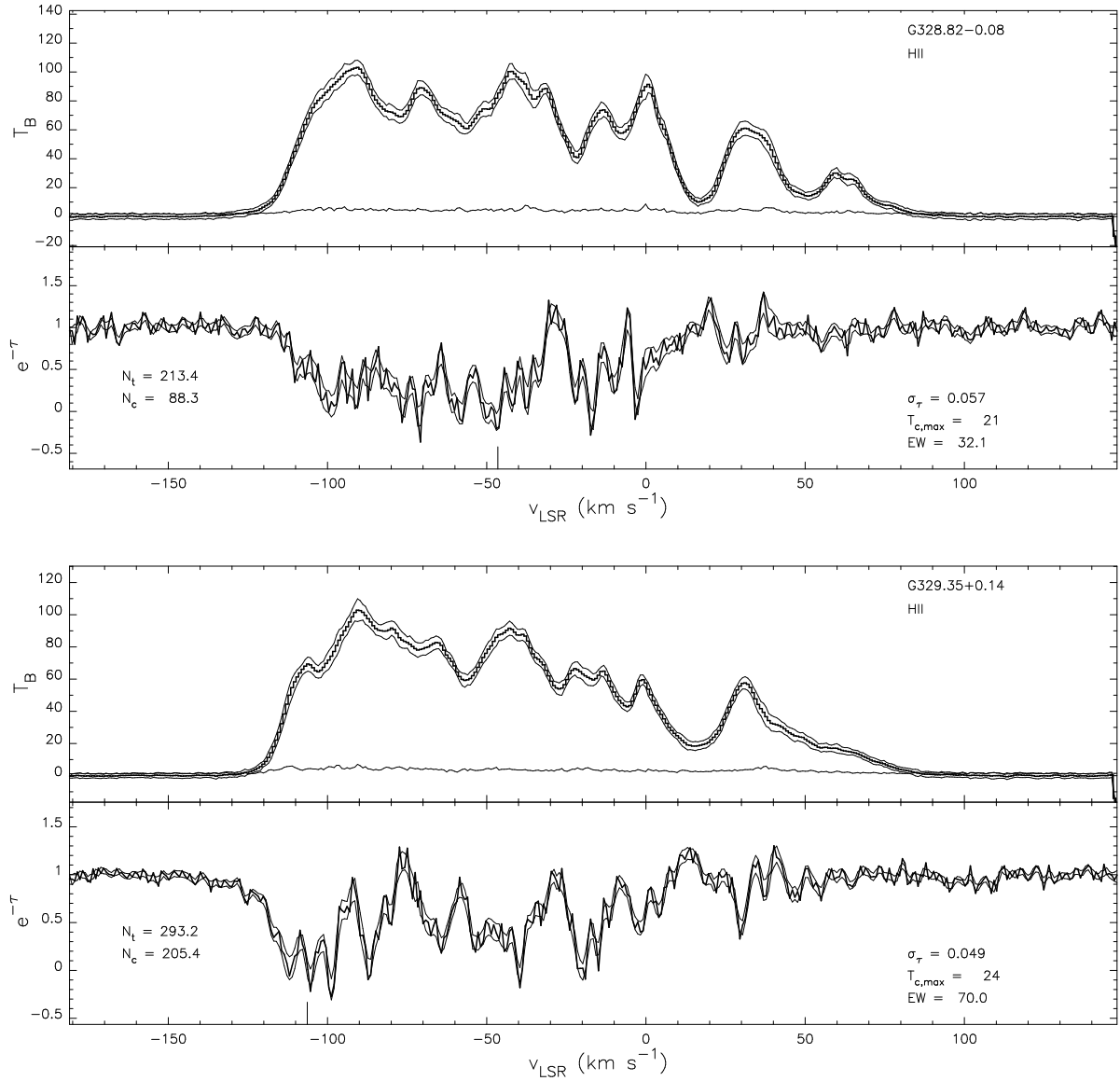}
\caption{Emission-absorption spectrum pairs for G328.82-0.08 and G329.35+0.14,
presented as in figure 2.}
\end{figure}

\begin{figure}
\epsscale{0.6}
%\plotone{fig6.ps}
\vspace{7in}
\epsfig{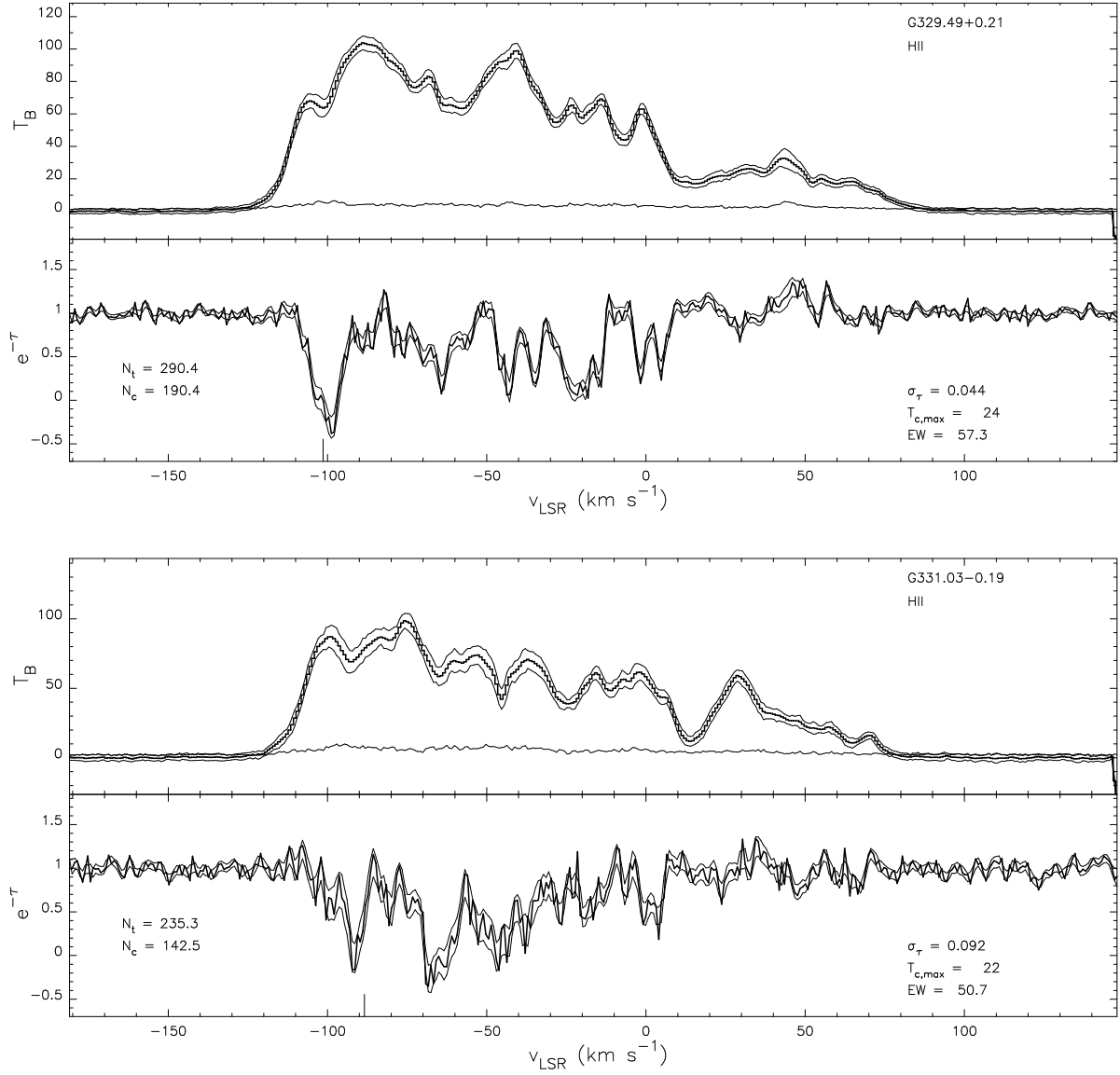}
\caption{Emission-absorption spectrum pairs for G329.49+0.21 and G331.03-0.19,
presented as in figure 2.}
\end{figure}

\begin{figure}
%\epsscale{0.6}
%\plotone{fig7.ps}
\vspace{7in}
\epsfig{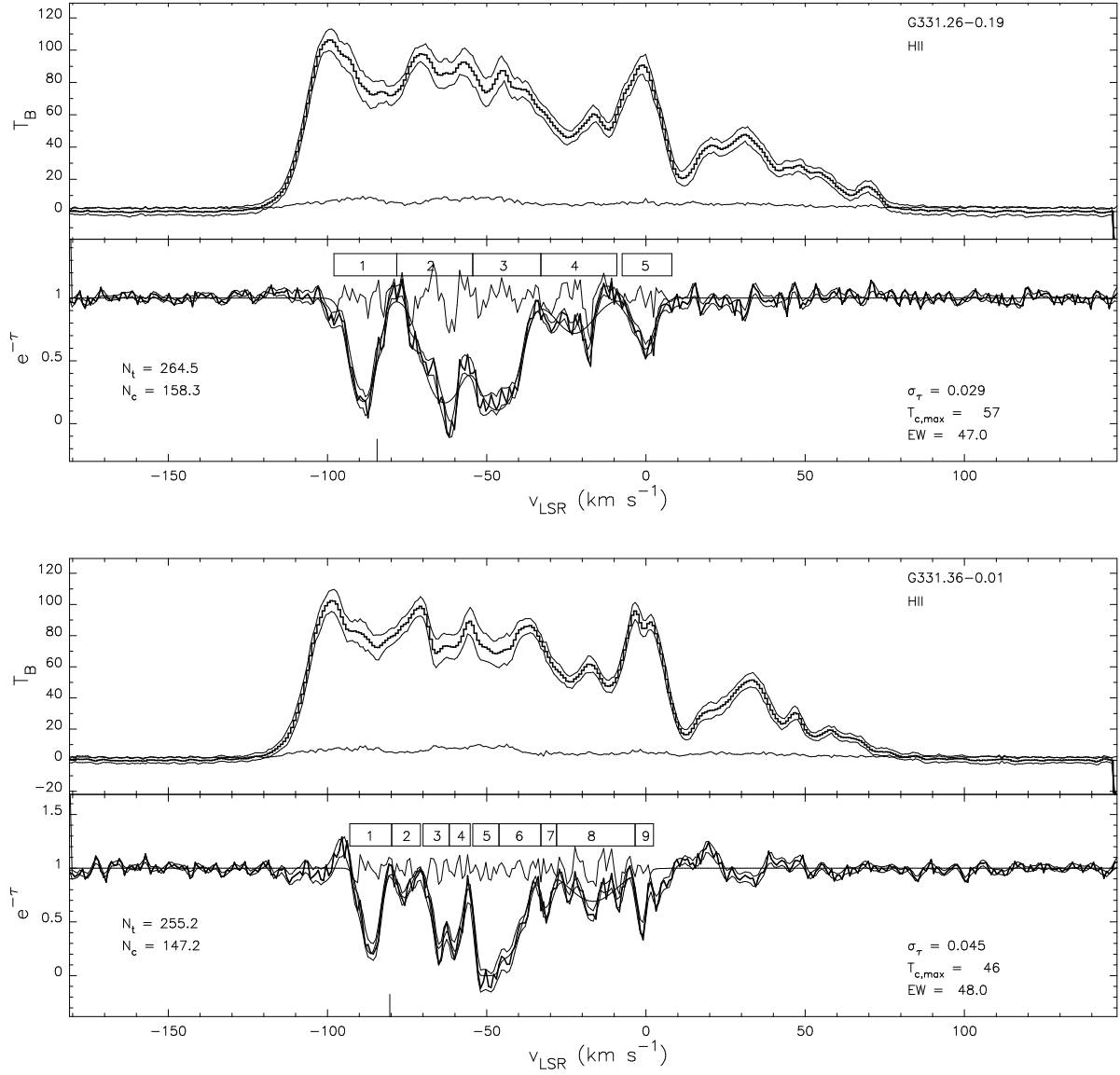}
\caption{Emission-absorption spectrum pairs for G331.26-0.19 and G331.36-0.01,
presented as in figure 2.}
\end{figure}

\begin{figure}
\epsscale{0.6}
%\plotone{fig8.ps}
\vspace{2.5in}
\epsfig{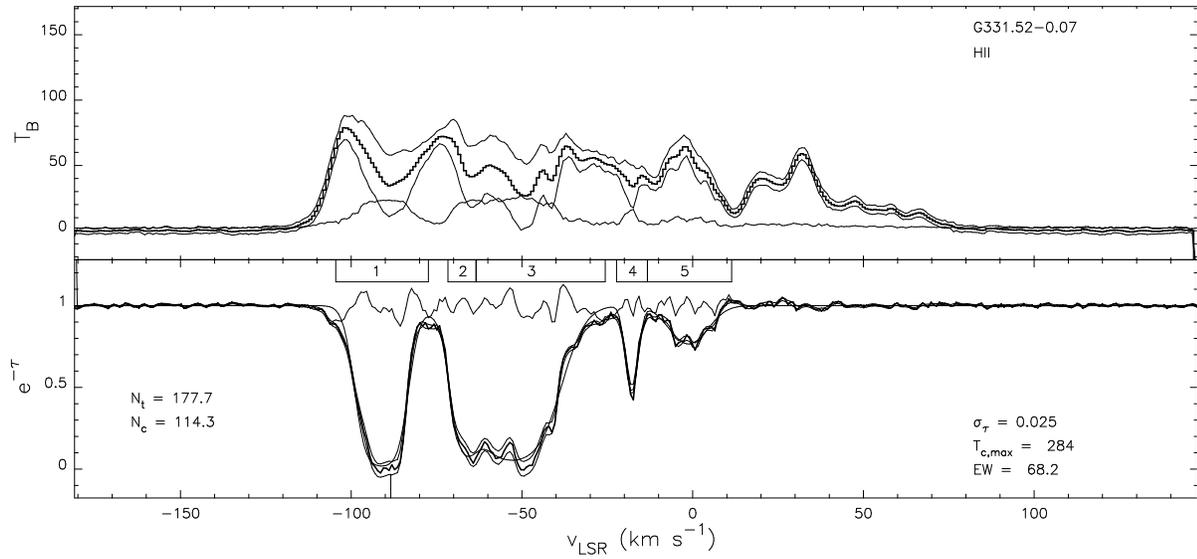}
\caption{Emission-absorption spectrum pair for G331.52-0.07,
presented as in figure 2.  As for G326.65+0.59 on figure 3,
this absorption spectrum was first observed by Goss et al. (1972),
with similar results to within their errors.}
\end{figure}

\section{Line Integrals and $\kappa (r)$}

The absorption spectra toward the weaker continuum sources have
so much noise due to emission fluctuations that we cannot 
compare the emission and absorption spectra channel-for-channel
to derive spin temperatures for each cloud or line component,
as we do for the spectra toward the stronger sources in section 4
below.  But we can work with the velocity integrals even for the 
noisier absorption spectra, since emission fluctuations produce
positive and negative ``pseudo-absorption'' (Radhakrishnan et al.
1972c) with no bias in the mean.  The velocity integral of the
optical depth spectrum is the equivalent width, $EW$,

\begin{equation}
EW \ = \ \int_{\Delta V}{ \tau(v)\ dv} 
\end{equation}

\noindent where the range of integration, $\Delta V$, is adjusted to
cover a specific interval along the line of sight using the
rotation curve.  Here we use the rotation curve of
Fich, Blitz, and Stark, (1989) :

\begin{equation}
R_{gal} \ = \ \frac{1.00746 R_{\odot} \Omega_{\odot}}{\Omega \ + \ 
0.017112 \Omega_{\odot}} 
\end{equation}

\noindent where $\Omega$ is the angular velocity of rotation around the Galactic
center, $R_{gal}$ is the Galactocentric
radius, and $\Omega_{\odot}$ and $R_{\odot}$ are the corresponding
values for the LSR at the solar circle (220 km s$^{-1}$ and 8.5 kpc). 

Kinematic distances have a near-far ambiguity symmetric
about the tangent point, but $R_{gal}$ is unique, so we can assign to each
radial interval in the inner galaxy a corresponding velocity interval,
$\Delta V$, and compute the corresponding EW.  Dividing each EW by the 
line of sight length, $\Delta s$, through the
corresponding annulus gives the mean opacity, $<\kappa>$, by

\begin{equation}
 \left< \kappa \right> \ \equiv \ \frac{EW}{\Delta s} \ = \ C_H^{-1}
\left< \frac{n}{T_{sp}} \right>
\end{equation}

\noindent where n is the H{\tt I} density and $T_{sp}$ the excitation temperature
of the $\lambda$21-cm line, and the line of sight average denoted by
the brackets, $\left< \ \right>$, is  
taken over $\Delta s$ which corresponds to the range of integration
in velocity, $\Delta V$, used to compute $EW$.  
$C_H$ is the familiar conversion constant $1.823 \times 10^{18}
\frac{cm^{-2}}{K \ km \ s^{-1}}$ which converts to
$5.9\times 10^{-4} \frac{cm^{-3} \ kpc}{K \ km \ s^{-1}}$, giving $\kappa$ 
units of km s$^{-1}$ kpc$^{-1}$.  In these units $\kappa$ typically
has values of a few to a few tens in the Milky Way disk.  Note that 
$<\kappa>$ is the frequency integral of the more familiar $\kappa_{\nu}$,
(e.g. Spitzer 1977 eq. 3-2 and 3-14), which has units (length)$^{-1}$,
although the range of integration may be only one spectral line channel.
\clearpage

\begin{figure}
%\epsscale{0.6}
%\plotone{fig9.ps}
\vspace{4in}
\epsfig{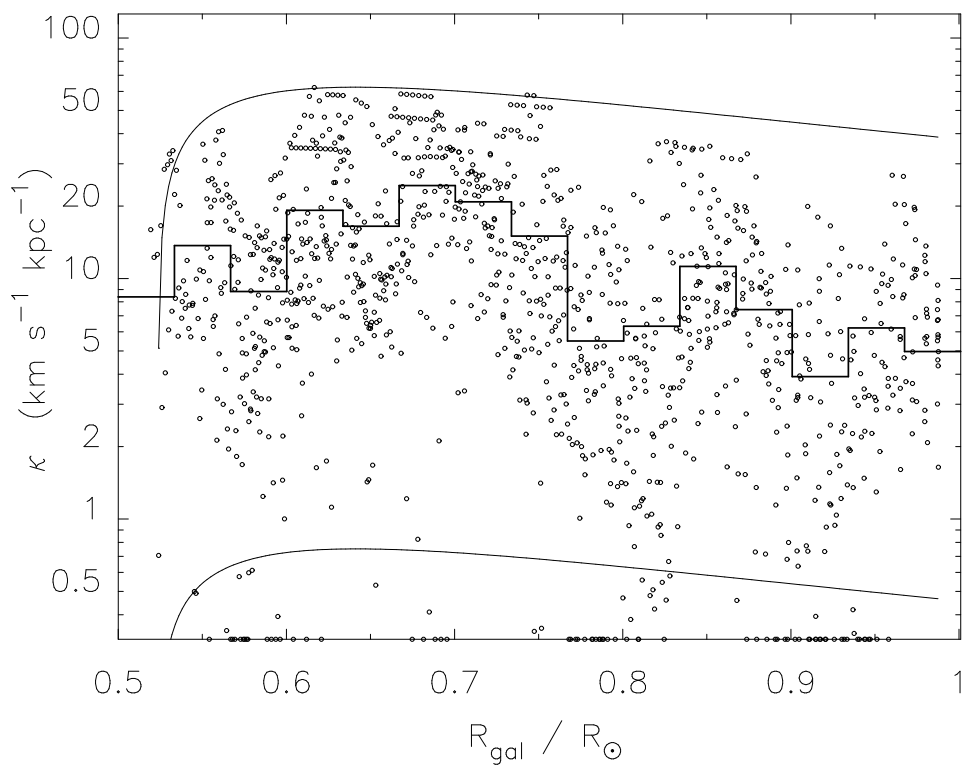}
\caption{Measurements of the 21-cm line opacity as a function
of Galactic radius.  The circles represent individual spectral 
channels, translated into optical depth per unit path length
using the velocity gradient, as described in the text.  The
histogram shows the averages of many channels on many lines
of sight for radial bins, weighting by line of sight length.
The upper and lower curves indicate the observational limits
due to saturation (top) and noise (bottom).}
\end{figure}

Results for $\left< \kappa \right>$ obtained by dividing the inner galaxy
into radial bins of width $\frac{1}{30} R_{\odot}$ are shown on figure 9.
The histogram shows an arithmetic average over the various
lines of sight through each annulus.
%We do not weight by the noise, $\sigma_{\tau}$ in this average, but
%we use only the directions toward continuum sources brighter than
%20K, which give $\sigma_{\tau} < 0.05$.  
%  program plotbye6 makes tables 1 and 2, as well as figures 2-8 !!
%sep02  The values plotted in
%sep02  this histogram are given on table 2, column 3, with the corresponding
%sep02  annulus edges, R$_{min}$ and R$_{max}$, given in columns 1 and 2, and
%sep02  the number of lines of sight passing through each annulus given in
%sep02  column 4.  
Also shown on figure 9 as small circles are
the individual channel values of $\tau$ in the absorption spectra,
converted to $\kappa$ by multiplying by the velocity gradient, 
$\frac{dv}{ds}$, given by the rotation curve (see Burton, 1988, for
a review of the significance of the velocity gradient).  
%Channels in the
%spectra toward the stronger sources, which contribute to the
%averages on table 2, are indicated by small circles, while 
%channels from spectra with higher noise are plotted as dots on figure 9.

The curves on figure 9 indicate observational selection based
on our noise level in $\tau$ for the case of G328.42+0.22, one of our
brighter sources.  The lower curve shows the opacity which would
result from optical depth equal to one sigma, as defined by the
emission fluctuations discussed above.  Weaker absorption than
this is not detectable, so multiplying by the velocity gradient
gives the corresponding lower limit for detectable opacity.  Spectra
toward fainter sources that give higher $\sigma_{\tau}$ will have
higher lower limits.  Points near or below this curve on figure 9
are upper limits on $\kappa$.  On the high side, when the absorption lines 
are very deep, the noise prevents us from distinguishing between
optical depths greater than
$\tau_{max}=-\ln \left( 2 \sigma_{\tau} \right)$.  The line
saturates at this point (typically $\tau \sim 3$), and we set the
optical depth to this value if $e^{-\tau}< 2 \sigma_{\tau}$.  Multiplied
by the velocity gradient, this maximum detectable $\tau$
gives an upper limit on the measurable $\kappa$, shown for the
case of G328.42+0.22 by the upper curve.  Again, for spectra with
higher noise level the upper limit is lower, as evidenced by 
chains of points at various levels corresponding to saturated lines.
Points near the upper curve on figure 9 are thus lower limits
on $\kappa$.

The velocity integrals of the absorption spectra are given on
table 1, columns 6 and 7.  Column 6 gives the integral over
the entire negative velocity range corresponding to the inner
galaxy.  Column 7 gives the integral over the restricted 
velocity range corresponding to the points on figure 9,
i.e. from zero km s$^{-1}$ to the recombination line velocity.
Whether the H{\tt II} region is at the near or far distance, we
can be confident that at least over this velocity range only 
the near-side gas can contribute to the absorption. 
For the H{\tt II} regions we stop 3 km s$^{-1}$ short of the 
recombination line velocity, to avoid the deep absorption usually
seen just beyond the H{\tt II} region's velocity.  The SNR
G328.42+0.22, is known to be beyond the solar circle on the far
side of the Galaxy (Gaensler, Dickel, and Green, 2000); in
this case we carry the velocity integration to the terminal velocity
at this longitude.  (Varying the 3 km s$^{-1}$ offset to
10 or even 15 km s$^{-1}$
has minimal effect on the opacities on figure 9,
%sep02  and table 2, 
although it reduces the number of lines of sight contributing 
to some of the annuli.)  
%The velocity ranges used for the 
%calculation of $EW$ and $\kappa$ are indicated on figures 2 - 8
%by the horizontal lines below the absorption spectra.

\begin{figure}
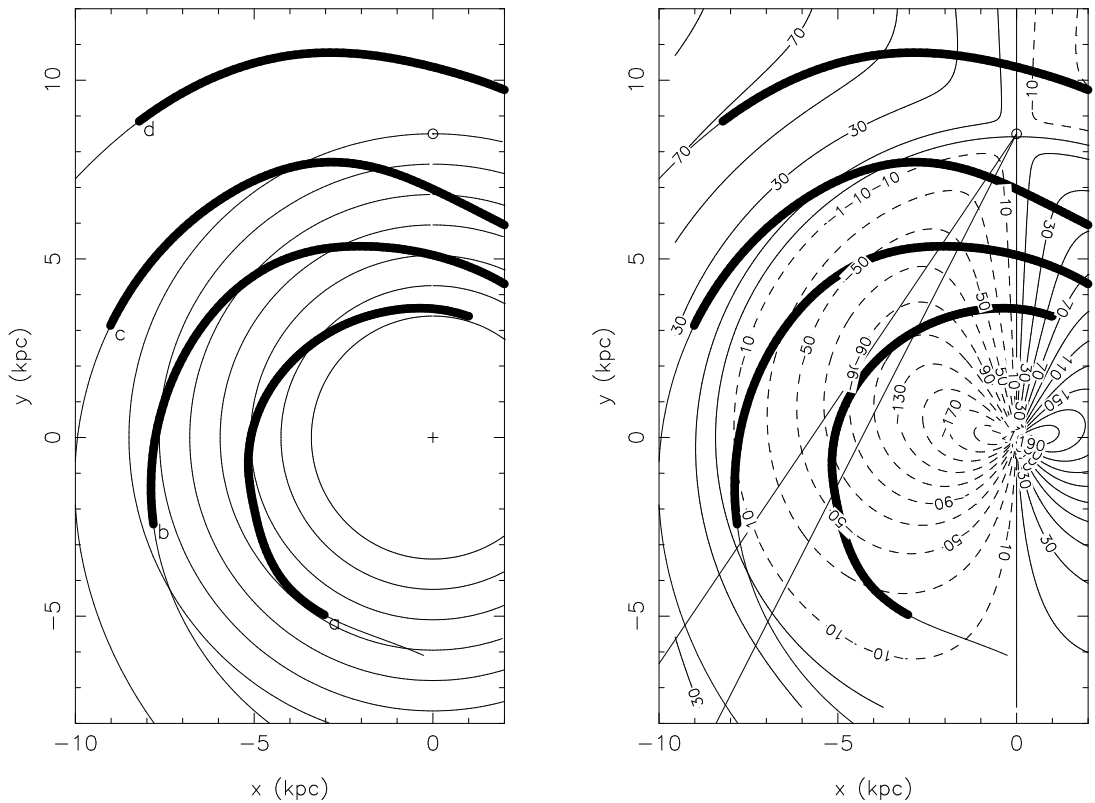

\epsscale{0.6}
%\plotone{fig10a.ps}
%\plotone{fig10b.ps}
\vspace{5in}
\epsfig{file=fig10a.ps,height=3in,angle=-90,
bbllx=8in,bblly=2.5in,bburx=4in,bbury=7.5in}
\epsfig{file=fig10b.ps,height=3in,angle=-90,
bbllx=8in,bblly=2.5in,bburx=4in,bbury=7.5in}
\caption{A face-on view of the Galaxy, with the longitude range
of the SGPS test region indicated by two lines at longitudes 326\arcdeg
and 333\arcdeg.  The right panel has the radial velocities predicted
from the rotation curve drawn as contours; the left panel shows
some annular bins which are crossed by the lines of sight at the
edges of the region.  The annuli used for the averages shown
on figure 9 are much narrower than those shown here, and there 
are many more of them.  Both figures have the spiral arm model
of Taylor and Cordes (1993) superposed.  Common names for the
spiral features are Norma (labelled ``a''), Scutum-Crux (``b''),
Sagitarius-Carina (``c''), and Perseus (``d'').}
\end{figure}

Figure 10 shows the geometry of the Galactic plane in the
fourth quadrant (from McClure-Griffiths et al. 2001).  The
sun's location is assumed to be at $(x,y)=(0,8.5)$, and we plot
lines of sight at longitude 326\arcdeg and 333\arcdeg, the boundaries of the
test region studied in this paper.  Annuli are plotted with 
radii from 0.4 to 1.0 $R_{\odot}$, and the major spiral features
from the model of Taylor and Cordes (1993) are reproduced.  We
see that the Norma spiral arm dominates the radial range 0.5 
to 0.6 $R_{\odot}$ near the tangent point, the Scutum-Crux arm
at the near distance dominates radii 0.6 to 0.7 $R_{\odot}$,
and Sagitarius-Carina dominates at about 0.85 $R_{\odot}$. 
There is a gap in the range 0.7 to 0.85 which is between arms.
This appears to be reflected in a dip in $\kappa$ at about 
0.8 $R_{\odot}$ on figure 9.  Even in this dip, and at all other
radii in the inner Galaxy, the opacity is larger than its solar
circle value of about 5 km s$^{-1}$ kpc$^{-1}$. 

This narrow longitude range may not be representative of the entire
inner galaxy, so we must wait for the full SGPS survey area to be
studied before deciding conclusively that the opacity increases
as we go to smaller Galactic radii.  In the first quadrant, Garwood
%sep02 and Dickey (1988) found about the same result for the solar circle
and Dickey (1989) found about the same result for the solar circle
opacity, but smaller values for the inner galaxy.  However,
recently Kolpak et al. (2002) have made more extensive 
observations with the VLA in the first quadrant, and they find
a trend similar to what we see on figure 9.  The difference
may be due to the small numbers of lines of sight sampled in the 
Garwood and Dickey study, which may have preferentially sampled
the regions between the spiral arms, which appear here to show
opacities of 5 km s$^{-1}$ kpc$^{-1}$ or less, in contrast to
the arm regions which can have $<\kappa> > 20$ km s$^{-1}$ kpc$^{-1}$.

%sep02  added this paragraph
If the increase in $<\kappa>$ by a factor of three to 
five  with decreasing Galactic radius from the solar circle to
the molecular ring region at about 0.5 $R_{\odot}$ is typical 
of the disk overall, we may ask whether it is due simply to an
increase in the surface density of gas.  The H{\tt I} surface
density as traced by 21-cm emission surveys
does not show an increase with decreasing R$_{gal}$
in this longitude range
(Burton 1988, fig. 7.15), but the molecular gas does.  The strong
variation of $<\kappa>$ tells us that the cool atomic gas, in
contrast to the warm neutral medium that dominates the
$\lambda$21-cm emissivity, is much more abundant in the inner
galaxy than at the solar circle.  This underlines the role of
the CNM clouds as an intermediate population
between structures in the WNM and the much denser
and colder molecular clouds.  
%sep02   end of added text

The integrals of the emission spectra are given on table 1, 
columns 8 and 9 in units of 10$^{20}$ cm$^{-2}$.  Column 8
gives the integral of the emission uncorrected for absorption, i.e. 

\begin{equation}
N_t \ = \ C_H \ \int{T_B(v)\ dv} .
\end{equation}

\noindent where we integrate over the full negative velocity range
corresponding to the inner Galaxy.    
In column 9 we take the emission integral over the restricted
velocity range $\Delta V$ corresponding 
to the absorption integral in column 7 (i.e. from zero to just
short of the recombination line velocity).  
For this velocity range we can correct
the values of emission for absorption to get a better estimate
of the true column density assuming a one-phase medium (see next
section).  That correction is given by 

\begin{equation}
N_c \ = \ C_H \ \int_{\Delta V}{T_B(v)\ \frac{\tau}{1\ - \ e^{-\tau}}\  dv} .
\end{equation}

\noindent The effect of this correction is to increase the column
density by a factor that is typically between one and two.  Values
of the ratio $f_c \ \equiv \ \frac{N_c}{N_t}$ {\bf for the same range of 
integration, $\Delta V$} are given on table 1, column 10.  
These correction factors show typical values of 1.4 to 1.6, which
is in agreement with similar correction factors found by Dickey and
Benson (1982).  Note that the $N_t$ values on Table 1, column 8
are typically larger than $N_c$ on column 9 because the range of
integration is broader for the former than for the latter, the
full inner Galaxy range for the former, and only to the recombination
line velocity for the latter.

Finally we can combine the velocity integrals of the emission and
absorption to get a velocity averaged spin temperature, 

\begin{equation}
\left< T_{sp} \right> \ = \ \frac{N_t}{EW \ C_H}
\end{equation}

\noindent which is the nominal value of the spin temperature
needed to give the emission and absorption integrals if the
gas were at a single temperature.  This is given on
table 1, column 11 for the velocity range corresponding
to columns 7 and 9.  Comparison of the emission
and absorption spectra channel-by-channel shows that this
value is far from realistic, even under the ``one-phase'' assumption.

\section{How to Combine the Emission and Absorption Spectra to find the
Diffuse Cloud Temperatures}

Methods for
combining the emission and absorption spectra to find the distribution
of spin temperatures in the interstellar H{\tt I} can get complicated.
The complications arise from blending of gas at different temperatures
in the same spectral channel.  Any line of sight at low Galactic latitudes
will contain a mixture of several cool clouds which appear as Gaussian
line components in the absorption spectrum, plus warm gas which is hardly
visible in the absorption spectrum, but which contributes the bulk of
the emission.  The challenge is to combine the absorption and emission
spectra in a way which separates these two thermal phases, giving an estimate
of the temperature of the cool gas without the bias introduced by the
warm gas emission.  In this section we discuss and compare several
mathematical techniques for this.

\subsection{Method 1.  The One-Phase Model}

The simplest way to combine the emission and absorption spectra is to
ignore the blending and compute the excitation temperature for each
velocity channel naively assuming all the gas at a given velocity is at
the same temperature.  This gives the simple formula:

\begin{equation} T_{sp}(v)\ \ = \ \ \frac{T_B(v)}{1\ - \ e^{-\tau(v)}} \end{equation}

\noindent where $T_B$ is the brightness temperature of the
emission in velocity channel $v$, and $\tau(v)$ is the 
optical depth of the absorption in the corresponding velocity channel.
This gives the harmonic mean of the temperatures of the
various neutral atomic regions which contribute to that velocity
channel, i.e. 

\begin{equation} T_{sp} \ \ = \ \ \frac{\sum_i{N_i}}{\sum_i{ \left(
\frac{N_i}{T_i} \right)}}\end{equation}

\noindent where $i$ is an index for the different regions on the line of sight
contributing, each with column density $N_i$ and excitation temperature $T_i$.
The distribution of $T_{sp}$ vs. number of channels for the nine sources
with $T_c >$30K is given on figure 11.  
We must assume that the temperatures resulting from equation 7 shown on
figure 11 are strongly biased to higher values than the true cool H{\tt I}
temperature due to blending with the warm gas.

\begin{figure}
%\epsscale{0.6}
%\plotone{fig11-9srcs.ps}
\vspace{4in}
\epsfig{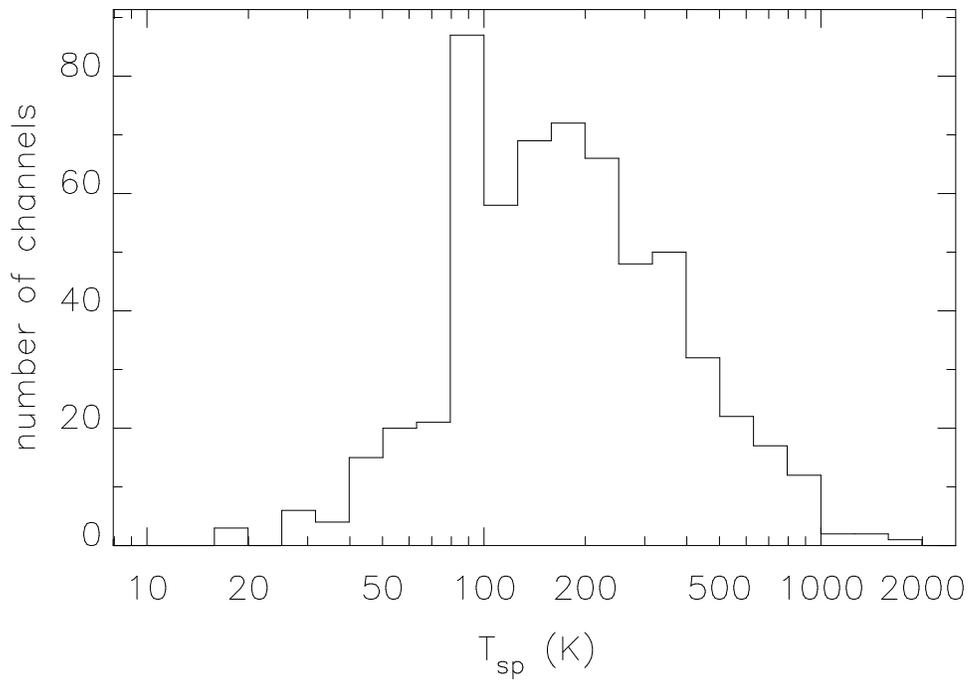}
\caption{Histogram of the temperatures derived from the naive, one
phase model ($T_{sp}$, eq. 7).  The weighting is simply by number
of channels, which roughly corresponds to weighting by volume rather
than by mass.}
\end{figure}

\subsection{Gaussian Fitting to the Absorption or to the Optical Depth}

     Improved estimates of the cool phase temperature are based on the shapes
of the features in the emission and absorption spectra.  A simple 
approach is to fit the absorption spectrum with a number of Gaussian
components.  Low latitude $\lambda$21-cm emission
spectra do not generally look like the
sum of distinct Gaussian components, but absorption spectra do, so 
this is a reasonable approach.  When the Gaussian components blend
together the best fit parameters become ambiguous, and for a 
complicated, low latitude spectrum even the number of Gaussian
components needed is not clear.  Thus there is no unique solution
to the Gaussian fitting problem.  
% deleted Sept 2002
%Even so, a Gaussian decomposition
%of the absorption spectrum is useful.  Figure 12 (top) shows a typical 
%Gaussian fit to a deep line in the G331.52-0.07 absorption spectrum, with the 
%residuals also shown.  

The fitted parameters
are obtained using an interactive
program based on the Levenberg-Marquardt method of non-linear
chi-squared minization (Press et al., 1992).
The numbered boxes above the spectra on figures 2-8 indicate
the velocity ranges dominated by the different line components
%sep02 (given on Table 3, columns 6 and 7).
(given on Table 2, columns 6 and 7).
Weaker components may be hidden below
the stronger ones in these velocity ranges.  Our detection limit
for absorption features is thus a function of velocity, and is
generally higher than that set by the noise.  
Note that these boxes do not represent velocity ranges over which
the fitting is performed.  In cases where several line components
are blended we perform the fitting of all Gaussian parameters
simultaneously over the full velocity range covered by all
components, plus some baseline on either side.

%\begin{figure}
%%\epsscale{0.6}
%%\plotone{fig12.ps}
%\vspace{3in}
%\epsfig{file=fig12.ps,height=3in,
%bbllx=0in,bblly=0in,bburx=4in,bbury=5in}
%\caption{Comparison of fitting ordinary Gaussians vs. saturated
%Gaussians (eq. 9) to the observed absorption spectrum
%($1 - e^{-\tau}$).  The flat top of the line is due to saturation,
%i.e. optical depth much greater than one.  An ordinary 
%Gaussian fit to $(1-e^{-\tau})$ overshoots the flattened peak,
%but still {\bf under}estimates the peak optical depth.}
%\end{figure}

For deep absorption lines such as those found at low latitudes,
it is preferable on physical grounds to fit the absorption 
spectrum with a sum of functions which are Gaussian in the
optical depth, $\tau$, rather than in the observed absorption,
$e^{-\tau}$.  We could do this
by fitting Gaussians to the log of the observed $e^{-\tau}$ spectrum,
but this amplifies the effect of the noise for the deeper lines; for
a deep line with noise we can get $(1-e^{-\tau})>1$ which causes
the log function to diverge.  It is better to do the fitting
using for the model a sum of functions of the form

\begin{equation}   
(1 \ - \ e^{-\tau(v)})_{model} \ = \  1\ - \ e^{- \sum_{i}{\ \left[
\tau_{0,i}\ e^{-\frac{(v-v_{0,i})^2}{2 \sigma_{i}^2}}\right] } }
\end{equation}

\noindent
where the parameters $\tau_{0,i}$, $v_{0,i}$, and $\sigma_{i}$ for 
each line component $i$ are varied to minimize $\chi^2$ in the
usual way (see table 2, columns 3-5).  In spite
%sep02 usual way (see table 3, columns 3-5).  In spite
of the double exponential this is no more
difficult to fit to the data than a sum of ordinary Gaussians, 
and it provides a better match to the flat tops seen on
the deeper lines.  More importantly, the values of
$\tau_{0,i}$ are significantly larger than the depths of the
corresponding Gaussian fits, because of saturation.
% changed sep 2002
Thus the derived column densities for the cool gas 
are larger, often by a factor of two or
more, than what would be obtained from simple Gaussian fits.
% deleted sep 2002
%Figure 12 illustrates the difference in the shape of the
%fitted function.
The fitted parameters are obtained using an interactive
program based on the Levenberg-Marquardt method of non-linear
chi-squared minization (Press et al., 1992).
The numbered boxes above the spectra on figures 2-8 indicate
the velocity ranges dominated by the different line components
%sep02 (given on Table 3, columns 6 and 7).
(given on Table 2, columns 6 and 7).  Weaker components may be hidden below
the stronger ones in these velocity ranges.  Our detection limit
for absorption features is thus a function of velocity, and is
generally higher than that set by the noise.  
Note that these boxes do not represent velocity ranges over which
the fitting is performed.  In cases where several line components
are blended we perform the fitting of all Gaussian parameters
simultaneously over the full velocity range covered by all
components, plus some baseline on either side.

We cannot estimate the cool phase temperature directly from
the widths of the Gaussians, however they are fitted, because
turbulence and random motions on all scales have the effect
of widening the line significantly beyond the thermal width
given by the Maxwellian distribution of atomic velocities.
So we need another approach to combine the emission and 
absorption spectra to get estimates for the temperatures
of the cool phase gas and undo the blending of equation 8.
The next simplest thing to do after the one-phase assumption
of equation 7 is to assume the H{\tt I} comes in two phases. 

\subsection{Method 2. Two Phase Linear Least-Squares Fitting}

The two phase assumption attributes all the absorption to
the cool clouds, while the emission spectrum includes the
cloud emission plus emission from a warmer medium which 
is necessarily more broadly distributed in velocity.  No 
assumption is made here about the spatial distribution
of either phase, but the fact that optical depth spectra can
be fitted by a sum of Gaussians suggests that the cool gas
is confined to discrete structures.  The assumption of
two phases means that the observed brightness temperature is 
%sep02  wording changed in sentence
given by

\begin{equation}
%sep02  B changed to obs
T_{obs} (v) \ \ = \ \ T_{w,f} \ + \ T_{w,b} \  e^{- \tau} \ + 
 \  T_{cool}\  \left( 1 \ - \ e^{- \tau} \right) \ + \ T_{cont} 
  \ e^{- \tau}  \end{equation}

\noindent
where the absorption comes from a cloud with temperature $T_{cool}$
and optical depth $\tau$, and
the warm, optically thin gas is partly in front
of the cloud, $T_{w,f}$, and partly behind it, $T_{w,b}$.  The 
last term provides for the presence of continuum emission, $T_{cont}$,
which is also behind the cloud, and hence is partially
absorbed by it.  Note that the contribution to the brightness
temperature from the warm gas has nothing to do with its 
physical temperature; so long as it is optically thin
its brightness
temperature is proportional only to the column density, $N_w$, in the 
velocity channel width $\delta v$, as
$T_{w}  =  \frac{N_w}{\delta v \  C_H}$.

%
%\noindent
%for the foreground ($f$), and similarly for the background ($b$).  

When we subtract the continuum from the emission spectrum
to obtain the line brightness temperature, $T_B(v)$, as a function
of Doppler velocity, $v$, 
%sep02  T_B(v) defined here 
then the last two terms in equation 10 combine as 

\begin{equation} T_B (v) \ \ = \ \ T_{w,f} \ + \ T_{w,b} \  e^{- \tau} \ +
 \  \left( T_{cool} \ - \ T_{cont} \right)
\  \left( 1 \ - \ e^{- \tau} \right)  \end{equation}

\noindent
This simplifies further if we consider the observed quantity 
$\left( 1 \ - \ e^{- \tau} \right)$ to be an independent variable
that we rename $x$, so that

\begin{equation} T_B (v) \ \ = \ \ T_{w,f} \ +
\ T_{w,b} \left( 1 - x \right) \ + \ 
x  \left( T_{cool} \ - \ T_{cont} \right) \end{equation}

Next we make an assumption that is supported by observations at
high and intermediate latitudes that the warm gas is broadly distributed
in velocity relative to the widths of the cool clouds.  So
over the velocity range of a single absorption line component we
can approximate the warm gas contribution as a linear function of
velocity, i.e.  $T_{w,f}   =  a_0  +  a_1 v $ and 
$T_{w,b}   =  b_0  +  b_1 v $ or for the total 

\begin{equation} T_{w,f}\ + \ T_{w,b} \  = \ c_0 \ + \ c_1 v .  \end{equation}

\noindent
with $a_0$, $a_1$, $b_0$, $b_1$, $c_0$, and $c_1$ all constants.
Then we have simply

\begin{equation} T_B (v) \ \ = \ \ \left( c_0 \ + \ c_1 v \right) \left( 1 \ - \ 
\epsilon x\right) \ + \ c_2 x \end{equation}

\noindent
where $c_2$ is the cool cloud temperature reduced by the background
continuum, $c_2 = (T_{cool} - T_{cont})$ and $\epsilon \equiv \frac{T_{w,b}}
{T_{w,f} + T_{w,b}}$ is the fraction of the warm gas which is behind
the cloud ($0 \leq \epsilon \leq 1$).  Here we will assume $\epsilon$ is
constant across the velocity width of each absorption line
component, though in general it may be a function of velocity.
Note that the continuum brightness, $T_{cont}$, is not the brightness
temperature $T_c$ of the background source towards which we measure
the absorption, but the diffuse continuum in the directions of the
nearby pointings which give the interpolated emission spectrum.

Given the observed absorption and emission spectra, we can
perform a simple {\bf linear} least-squares fit
to the data in order to determine
the values of $c_0$, $c_1$, and $c_2$ in equation 14.  The independent
variables are $v$ and $x$, and the dependent variable, $T_B$, we will 
call $y$ for simplicity.  The solution is obtained by minimizing the
value of $\chi^2$ in the same way as for polynomial fitting,
which leads to a matrix inversion
problem where the elements of the matrix are moments of various 
combinations of $v$, $x$, and $y$.  For example, 

\begin{equation} \overline{vxy} \ =
\ \frac{1}{n} \ \sum_i \ v_i \cdot x_i \cdot y_i \end{equation}

\noindent
where the sum is taken over all spectral channels, $i=$ 1 to
$n$, covered by a
distinct absorption feature (which presumably corresponds to a single
cool phase temperature).  The parameter $\epsilon$ is carried through
all the calculations explicitly, so that results can be found for any
selected value of $\epsilon$.  The advantage of this approach, compared
with the Gaussian fits discussed above, is that there is no ambiguity in
the results; the unique solution is found directly with no iteration
and no need for a ``first guess'' input by hand.
The equations needed to find this best fit are given in the appendix.
%\clearpage

%sep02 \begin{figure}
%\epsscale{0.9}
%\plotone{fig13.ps}
%sep02 \vspace{6in}
%sep02 \epsfig{file=fig13.ps,height=4in,angle=-90,
%sep02 bbllx=8in,bblly=1.5in,bburx=4in,bbury=6.5in}
%sep02 \caption{Detailed analysis of the variation of brightness 
%sep02 temperature vs. optical depth.  The top left panel shows at
%sep02 the top the emission-absorption spectrum pair toward G326.45+0.90 
%sep02 as in figure 3.  The upper portions of the other panels 
%sep02 show the same two spectra, but blowing up the velocity ranges
%sep02 indicated by the numbered boxes.  The emission spectrum has
%sep02 superposed the least squares fit, and below the residuals (data minus
%sep02 fit).  The linear function, $T_w$, is also shown with the
%sep02 emission spectrum.  The graphs just below the absorption
%sep02 spectra are plots of $T_B$ {\it vs.} $x\equiv \left( 1 - e^{-\tau}
%sep02 \right)$.  The {\it x}'s on these plots show the individual
%sep02 channel values.  The solid curve on the $T_B$ {\it vs.} $x$
%sep02 plots shows the least squares fitted function of the form of
%sep02 equation 14.  The coefficients of this function are indicated
%sep02 for each absorption feature sub-plot by the equation at the
%sep02 bottom. } 

%sep02 \end{figure}

\begin{figure}

%\epsscale{0.9}
%\plotone{fig14.ps}
\vspace{6in}
\epsfig{file=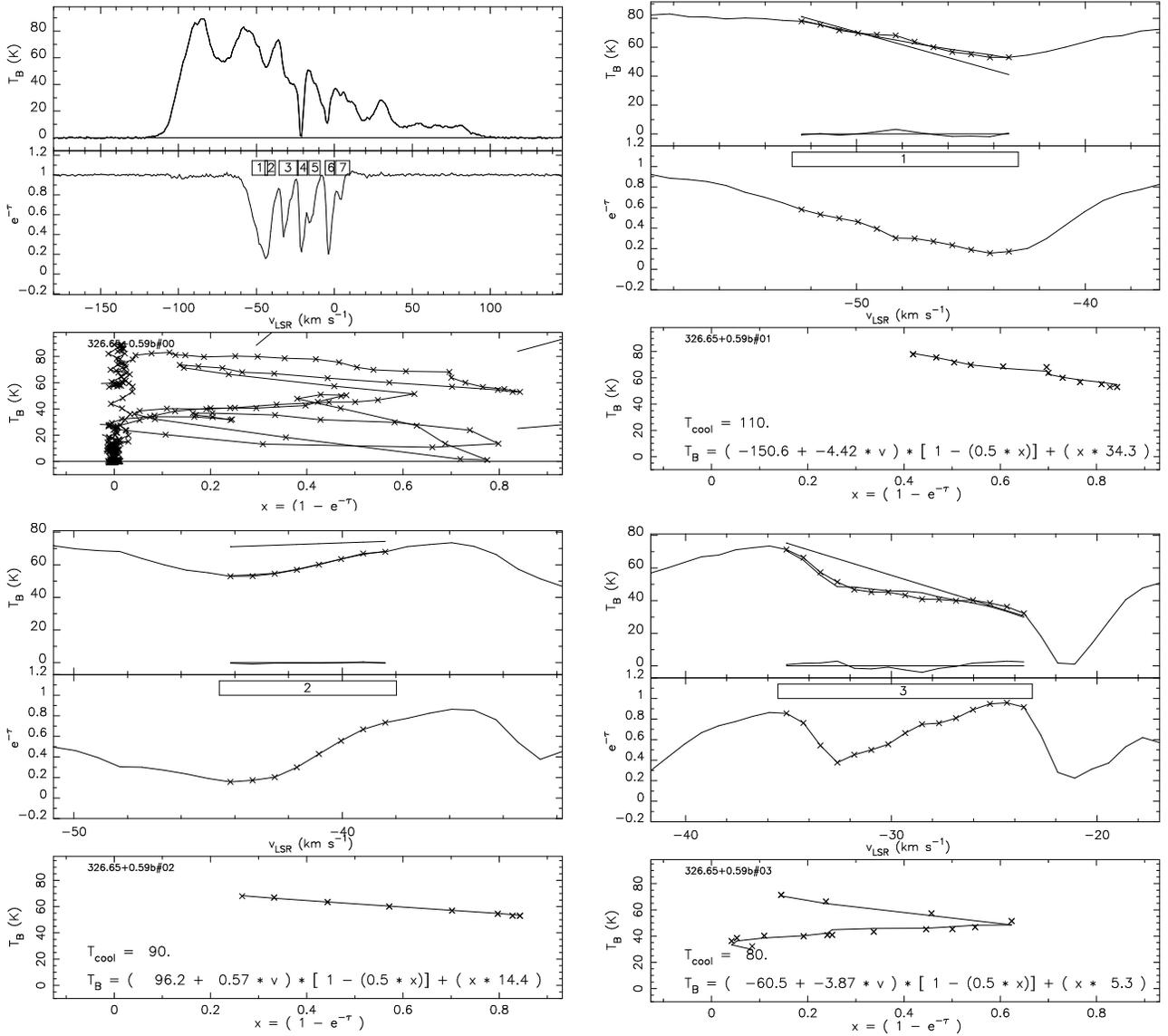,height=4in,angle=-90,
bbllx=8in,bblly=1.5in,bburx=4in,bbury=6.5in}
\caption{Detailed analysis of the variation of brightness
temperature vs. optical depth.  The top left panel shows at
the top the emission-absorption spectrum pair toward G326.65+0.59
as in figure 3.  The upper portions of the other panels
show the same two spectra, but blowing up the velocity ranges
indicated by the numbered boxes.  The emission spectrum has
superposed the least squares fit, and below the residuals (data minus
fit).  The linear function, $T_w$, is also shown with the
emission spectrum.  The graphs just below the absorption
spectra are plots of $T_B$ {\it vs.} $x\equiv \left( 1 - e^{-\tau}
\right)$.  The {\it x}'s on these plots show the individual
channel values.  The solid curve on the $T_B$ {\it vs.} $x$
plots shows the least squares fitted function of the form of
equation 14.  The coefficients of this function are indicated
for each absorption feature sub-plot by the equation at the
bottom. }

\end{figure}

\begin{figure}
%\epsscale{0.9}
%\plotone{fig15.ps}
\vspace{6in}
\epsfig{file=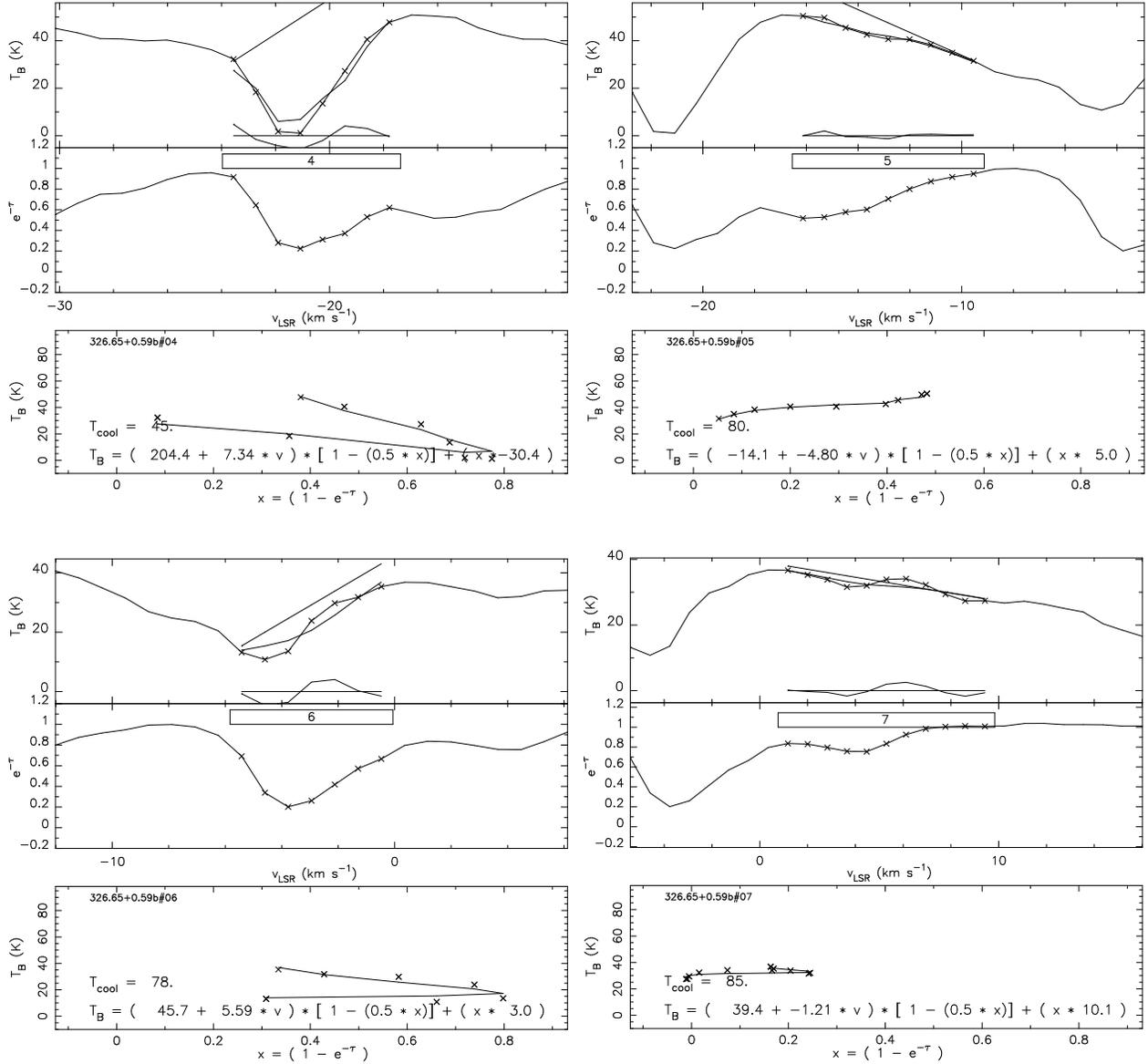,height=4in,angle=-90,
bbllx=8in,bblly=1.5in,bburx=4in,bbury=6.5in}
\caption{Plots of the spectral features toward G326.65+0.59,
continued from figure 12.}
%sep02 
\end{figure}

%This fitting method is illustrated in figure 13-14, which show plots of 
This fitting method is illustrated in figure 12-13, which show plots of 
%sep02 
the emission brightness temperature, $T_B$ against the observed
absorption, $1-e^{-\tau}$.  Different panels on figures 12-14 show  
%sep02  separate line components toward G326.45+0.90
separate line components toward 
G326.65+0.59, corresponding to the numbered
boxes on the upper panel of figure 3.  The solid curves show the fit results,
and the symbols show the measured values in each velocity channel.
In most cases the fits are quite good, considering that there are
only three free parameters, and typically 10 to 15 independent 
pairs of measured quantities.

The approach taken here is very similar to that of Mebold et al. (1997)
and Dickey et al. (2000) in their studies of the cool H{\tt I} in the 
Magellanic Clouds.  In those papers the fitting was done by hand
by determining the ``ridge-line'' of each feature measured on the
%$T_{em}$ vs. $\tau$ plane like those in figures 13 and 14.  That approach
$T_{em}$ vs. $\tau$ plane like those in figures 12 and 13.  That approach
works best for simple spectra without blended absorption lines, as
seen in the Magellanic Clouds and some high and intermediate latitude
Galactic directions.  At low latitudes the many overlapping absorption
features make it difficult to draw the ridge lines by hand, so the
fitting technique described in this section is needed.  The results
for unblended lines are very similar to those derived using the
ridgeline method.

\begin{figure}
%\epsscale{0.6}
%\plotone{fig16.ps}
\vspace{4in}
\epsfig{file=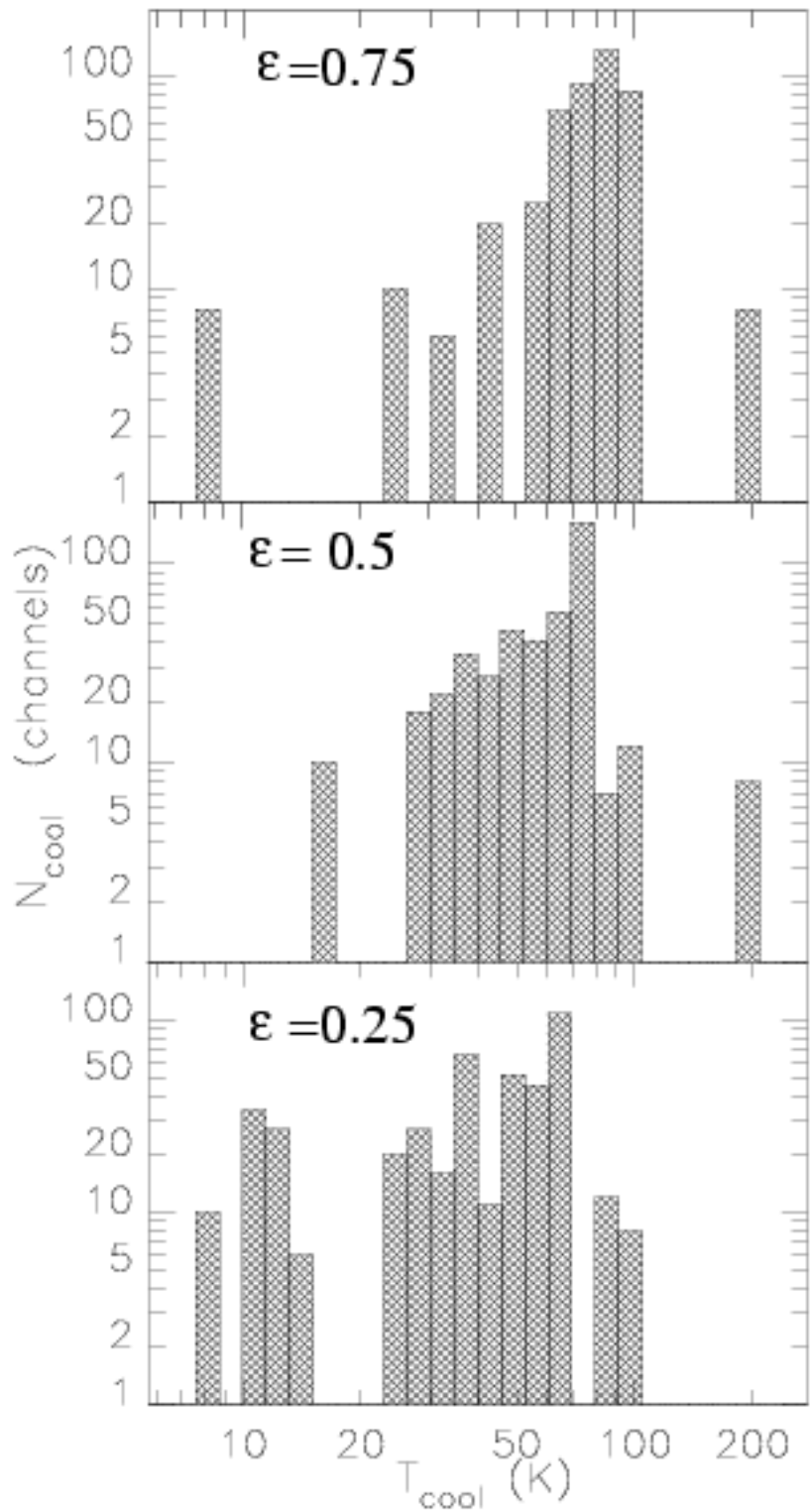,height=4in,
bbllx=3in,bblly=2in,bburx=-1in,bbury=6in}
\caption{Histogram of $T_{cool}$.  The histograms show
the number of spectral channels at each temperature, 
derived using method 2 (linear least squares).  Results
for three values of $\epsilon$ are shown.  The median
values of $T_{cool}$ are 79 K, 65 K, and 51 K for $\epsilon$ = 0.75,
0.5, and 0.25, respectively.}
\end{figure}

The fitted values of $c_0$, $c_1$, and $c_2$ tell us both the 
temperature of the cool cloud ($T_{cool}=c_2+T_{cont}$) and the
linear fit to the warm gas emission across the velocity range of
the absorption line component (equation 13).  These fit results
are shown on table 2, columns 8 through 10.  
For this analysis we use only the five spectra toward continuum
sources with T$_{c,max} > $75 K, for which the rms errors in the
absorption are less than 0.025.  From these fits we can
derive new histograms of cool gas temperatures.  These are
%shown on figure 15, for three values of $\epsilon$=0.25, 0.50,
shown on figure 14, for three values of $\epsilon$=0.25, 0.50,
and 0.75, which should bracket the
true foreground - background distribution. 

%sep02 are shown on table 3, columns 8 through 10.  
%sep02 Since the results
%sep02 depend on the value of $\epsilon$, we give three examples, 
%sep02 $\epsilon$=0.25, 0.50, and 0.75, which should bracket the
%sep02 true foreground - background distribution. 
%sep02 From these we can
%sep02 derive new histograms of cool gas temperatures, for value
%sep02 of $\epsilon$, as shown on figure 15.  
%sep02 

\begin{figure}
%\epsscale{0.6}
%\plotone{fig17.ps}
\vspace{6in}
\epsfig{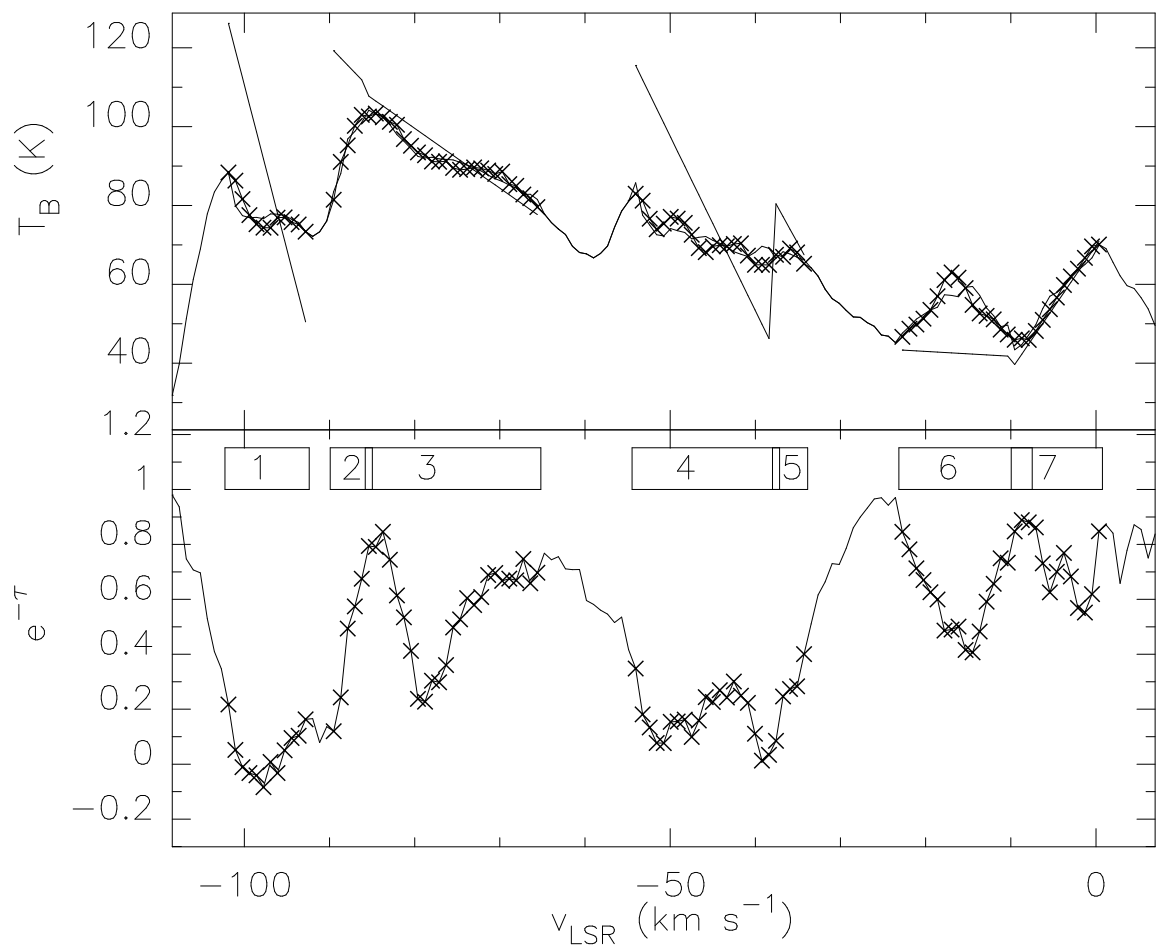}
\caption{Detail of the linear least squares fitting
results for G328.31+0.45.  The upper spectrum (emission)
shows both the data ({\it x}'s and the fitted emission
profile, which are nearly indistinguishable, as well as
the linear fits to the warm phase emission ($c_0 + c_1 v$).
These piecewise linear fits are not continuous, either with
each other or with the emission spectra on either side
of the absorption features.  This motivates the third
method (non-linear least squares) for finding $T_{cool}$,
which forces the warm phase models to be continuous.}
\end{figure}

%sep02 Figure 16 shows simply the number of spectral channels covered
Figure 14 shows simply the number of spectral channels covered
by clouds with a given value of $T_{cool}$. This is not 
a mass weighted average, nor a volume average, but just
the statistics of clouds sampled on the small number of lines
of sight toward the five strongest continuum sources.  
Since the sample is small, and groups of channels come 
out with the same cloud temperature, the histograms have
gaps at the lower temperatures, which are certainly unrealistic.
These histograms must be considered merely representative
of the distribution of H{\tt I} cloud temperatures, with the
general conclusion that most clouds have cool phase
temperatures in the range 30 to 100 K, with occasional
cooler values.  

The parameter $\epsilon$ is undetermined
for any given cloud.  In general we may expect roughly half of
the warm-phase gas to be in front of the absorbing cloud (at 
a given velocity) and half behind, which means $\epsilon \simeq 0.5$
is a good guess.  The lower panel, which shows the result from
setting $\epsilon = 0.25$ for all clouds, certainly underestimates
the temperatures of many clouds, since if only one quarter of
the warm phase emission comes from behind the cloud, the cloud's
optical depth must be much higher, and hence its temperature cooler,
then for $\epsilon$=0.5.
Similarly, the upper panel, which assumes $\epsilon = 0.75$ for
all clouds, must overestimate the temperature of most clouds.
An exception is the cloud centered on -5 km s$^{-1}$ toward G326.45+0.90.
Fitting equation 14 to this
results in a very cold temperature, 13 K assuming $\epsilon = 0.75$,
just 2 K if $\epsilon = 0.5$, and -9 K for $\epsilon = 0.25$.
%sep02 (In the middle and lower panels of figure 15 this cloud drops off
(In the middle and lower panels of figure 14 this cloud drops off
%sep02 
the left edge of the plot.)
The latter two are clearly unphysical, so in this case we may
conclude that $\epsilon$ must be larger than $\sim 0.6$, and the
cloud must be quite cold, strictly less than 25K, which is the
value resulting from assuming $\epsilon=1$.

\subsection{Method 3.  Two Phase Non-Linear Least Squares Fitting}

The linear least-squares fit technique described by equations 10-15
and in the appendix
does a good job of matching the observed values of $T_B$ using 
the measured absorption values as the independent variable, but this
technique ignores what we know about the warm gas emission at velocities
{\bf on either side} of the absorption line.  Ideally
the linear approximation to $T_w$
given by $c_0+c_1 v$ should be continuous with the emission spectrum 
above and below the velocities covered by the absorption line, and in 
the case of multiple overlapping absorption lines, piecewise linear
fits should be continuous between sequential velocity ranges.  But this
constraint is not incorporated into the fitting algebra, and typically
%sep02 it is not satisfied.  An example is shown on figure 16, where the
it is not satisfied.  An example is shown on figure 15, where the
%sep02 
separate least squares fits toward G328.31+0.45 are plotted together.
The discontinuities between
successive linear approximations to $T_w$ show that we are still missing 
some information that could improve the model of the H{\tt I} temperatures.

To force continuity in the $T_w$ model we have to return to a non-linear
least-squares fitting technique, based again on the Levenberg-Marquardt
method.  This is a more complicated application, however, as we are now
fitting a model to $T_B(v)$ using two independent variables, $v$ and 
$x\equiv (1-e^{-\tau(v)})$, and a set of parameters.  The parameters
are $c_2 \equiv (T_{cool}-T_{cont})$ for each line component, and
now instead of $c_0$ and $c_1$ to describe the warm gas emission for
each velocity range, we have just one parameter for each contiguous
box. For distinct velocity ranges (non-blended lines) the 
warm phase emission is fully determined by its values on either
side of the absorption, and there is only one parameter to fit ($c_2$). 
For the case of overlapping lines, we can adjust the warm phase
component arbitrarily at the boundary between the boxes, but
in each box we stick to a linear interpolation for our estimate
of the warm phase emission.  Thus for blended absorption lines
we are fitting both the cool phase temperature and 
the value of $T_w$ at each velocity
edge where individual components overlap.  The piecewise
linear approximation to $T_w(v)$ is now fully specified by the 
values $T_w(v_i)$, where the set of $v_i$'s are the boundaries of the
%sep02 velocity ranges represented by the boxes on figure 16. 
velocity ranges represented by the boxes on figure 15. 
%sep02 
This forces
the piecewise linear model $T_w$ to be continuous from one component
to the next.  If we further require that $T_w$ be continuous with 
$T_B$ on the first and last velocity edge, i.e. where the absorption
drops below its detection threshold, then for a set of $n$ overlapping
line components we have to fit just $n$ values of $c_2$ and $(n-1)$
values of $T_w(v_i)$, where $i=2,...,n$ and $T_w(v_1)$ and $T_w(v_{n+1})$
%sep02 are fixed.  Figure 18 shows the same
are fixed.  Figure 16 shows the same
spectral pair as in figure 15 (G328.32+0.45), but now fitted
%sep02 
using method 3.  The estimate of the warm phase contribution
is much better behaved.  The residuals of the fit to $T_{em}$
are shown at the bottom.  Given that there are fewer than two free
parameters to fit for each absorption component, the agreement
between the data and the fit is impressive.

Figure 17 (lower panel) illustrates the distribution of cool 
%sep02 
phase temperatures resulting from this fitting for $\epsilon$=0.5,
compared with a copy of the histogram for $\epsilon$=0.5 from
figure 14.  The third method of fitting changes the results
%sep02 
slightly, mainly by reducing the higher values of $T_{cool}$
and so narrowing the peak near 70K.  There is still a tail
reaching to low temperatures, and the median value of $T_{cool}$
is hardly changed (67 K vs. 65 K from method 2).

\begin{figure}
%\epsscale{0.6}
%\plotone{fig18.ps}
\vspace{6in}
\epsfig{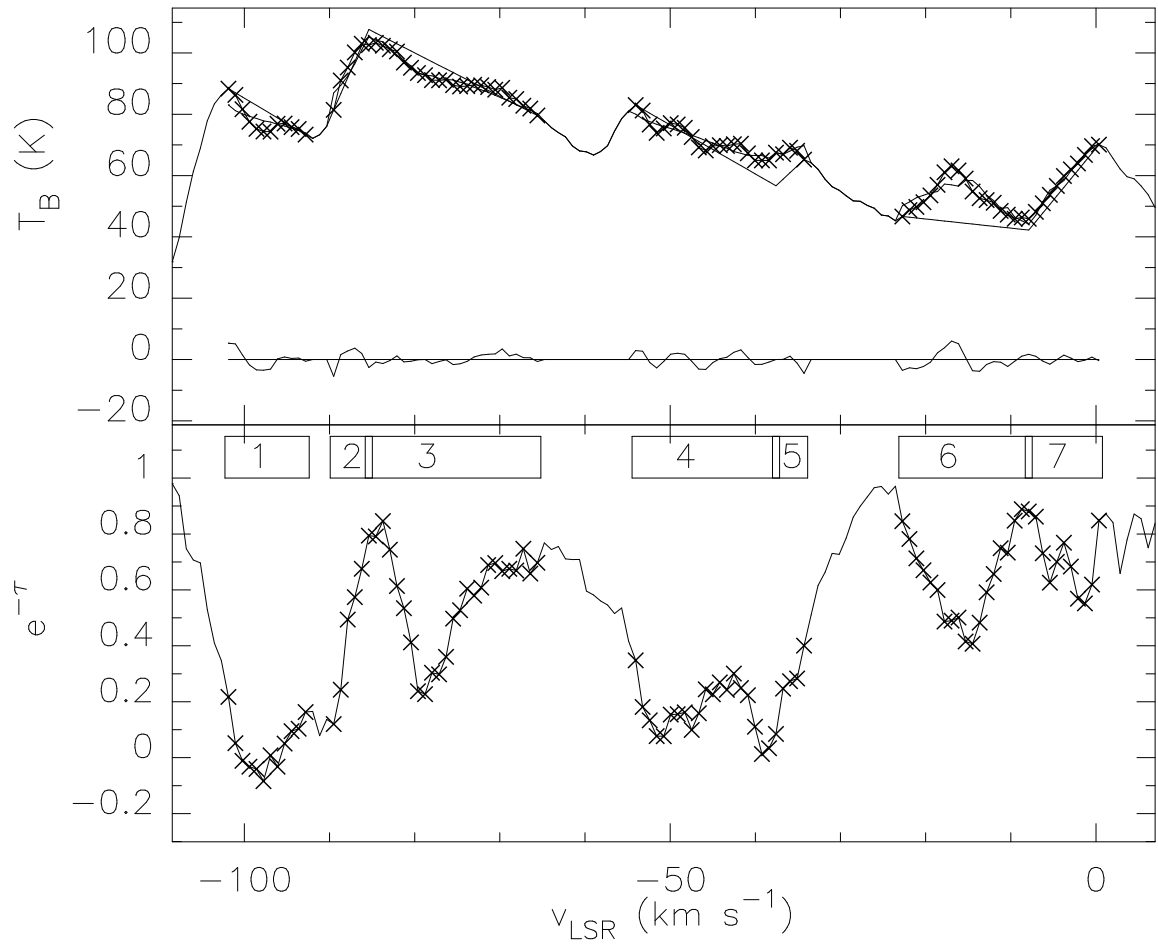}
\caption{Detail of the non-linear least squares fitting
%sep02 results for G328.31+0.45, as in figure 17. The piecewise
results for G328.31+0.45, as in figure 15. The piecewise
linear fits for the warm phase emission are now much more
reasonable.  The data and the fitted values for the total
emission are also shown, but almost indistinguishable.
The residuals (data minus fit) are shown at the bottom. }
\end{figure}

For a single, isolated absorption component, the warm gas level is determined
by a linear interpolation between the values of $T_B$ on the edges of
the line, so there is only one parameter to fit, i.e. $c_2$, that
determines the cool gas temperature.  This effectively reduces to the
method of interpolation used by Kalberla, Schwarz, and Goss (1985)
in their study of the
emission and absorption in the vicinity of 3C147, and by Mebold et
al. (1982) in their study of several high and intermediate latitude
regions.   
If we were to assume that the absorption and warm phase emission 
components could all be fitted by blended Gaussians then we could 
derive the cool phase temperature directly.  This is the approach
taken by Heiles and Troland (2002a,b) in their Arecibo survey
of absorption and emission at intermediate latitudes.
In appendix 2 we make a detailed comparison of their results
with the results of our Method 2, and find that the differences
are minor.

\begin{figure}
%\epsscale{0.6}
%\plotone{fig19.ps}
\vspace{4in}
\epsfig{file=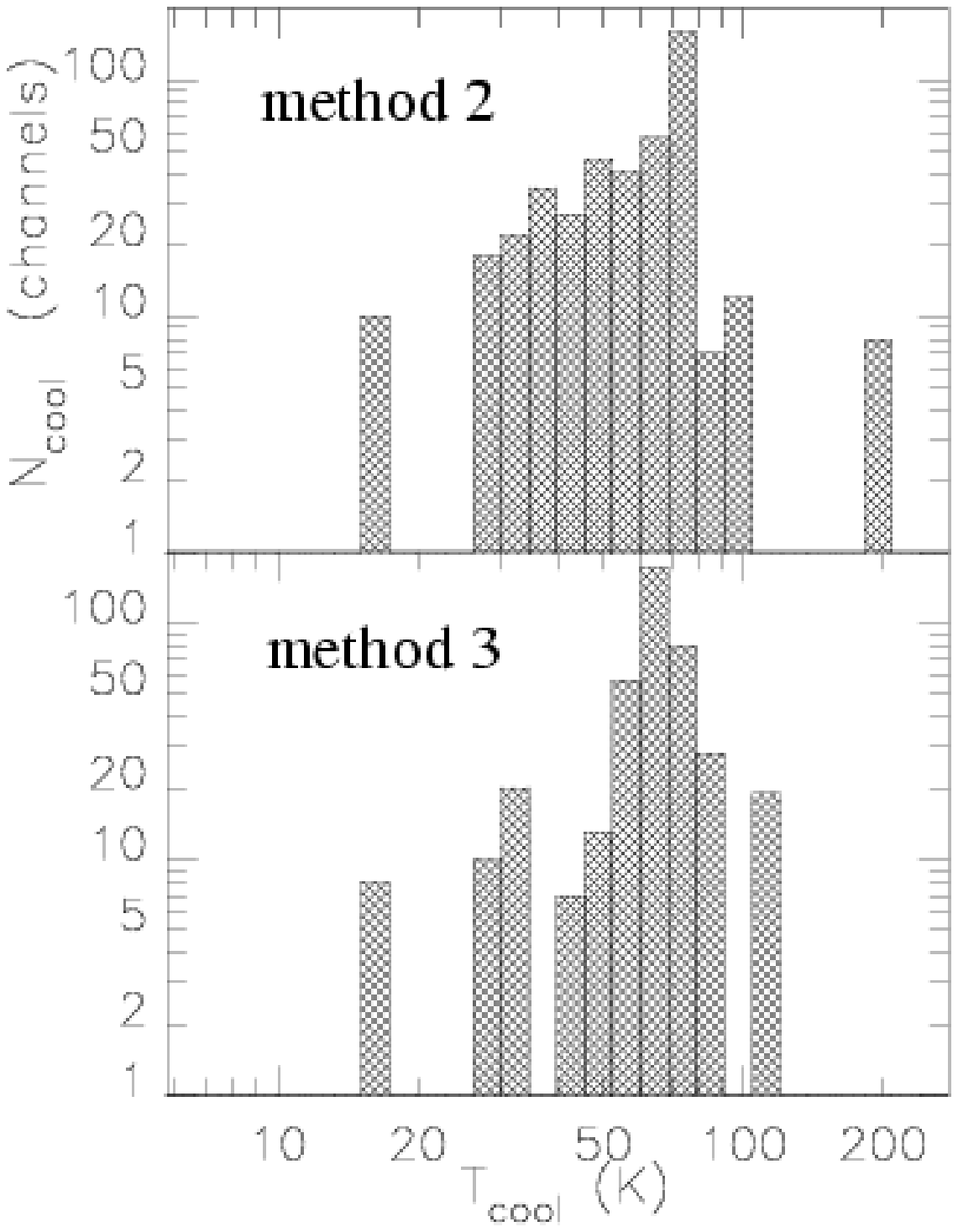,height=4in,
bbllx=3in,bblly=2in,bburx=-1in,bbury=6in}
\caption{Comparison of the $T_{cool}$ distributions
obtained with method 2 and method 3.  Both histograms
assume $\epsilon$=0.5.  The upper histogram gives
median $T_{cool}$ = 65 K, the lower histogram
gives 67 K.  Both are unweighted counts of 
spectral channels at each temperature, without
regard to the total mass of gas in each channel.}
\end{figure}

\section{Discussion}

\subsection{The Peak Brightness Temperature}

The mean opacity, $<\kappa>$, derived in section 3 above plays
a fundamental role in determining the radiative transfer of the
21-cm line across the Galactic plane.  
At low latitudes, cool clouds with optical depths of one or more
are common enough that some, but not all, of the velocity range
in a typical direction will be covered by optically thick gas.
How likely this is depends on a contest between  $<\kappa>$ and
the velocity gradient, $\frac{dv}{ds}$, along the line of sight.
These two quantities have the same units (km s$^{-1}$ kpc$^{-1}$),
so we can make a dimensionless ratio, $\rho$, defined by
%sep02 changed R to rho

\begin{equation}
\rho \ \ \equiv \ \ \frac{\kappa}{\left| \frac{dv}{ds} \right|}
\end{equation}
%sep02 changed R to rho

\noindent
that depends on the local properties of the medium (the abundance of
cool clouds) as well as on our particular vantage point (the longitude)
and the shape of the rotation curve.  In the extremely low velocity
gradient case, as near longitudes 0\arcdeg and 180\arcdeg where 
$\frac{dv}{ds} \ \simeq \ 0$, this ratio
$\rho$ is much greater than one, and the optical depth is high at
%sep02 changed R to rho
all allowed velocities.  In contrast, in the outer galaxy at a 
longitude where the velocity gradient is moderate, e.g. $\pm90\deg$
for which $\left| \frac{dv}{ds} \right| \sim$ 10 km s$^{-1}$ along
most of the line of sight,
high optical depths are rare, since $\rho$ is
%sep02 changed R to rho
less than one ($<\kappa>$ in the outer Galaxy is mostly below
5 km s$^{-1}$ kpc$^{-1}$).  Since the absorbing gas is not
smoothly distributed in space but is collected in discrete 
clouds with small filling factor, the spectra show occasional narrow 
features of significant optical depth even in regions where the
mean optical depth is low.  The ratio $\rho$ is useful for predicting
%sep02 changed R to rho
optical depths averaged over velocity
width $\Delta V$ broader than the linewidth of a single 
cloud, $\delta v \lesssim 5$ km s$^{-1}$.

Using the cool cloud temperature distribution, derived in section 4,
together with $<\kappa>$, we can derive an approximate value for the
peak brightness temperature of the Galactic 21-cm emission.  It was
noticed as early as the 1950's in surveys of the Galactic plane that
the 21-cm emission typically peaks at $T_B \sim $100 to 125 K,
regardless of the direction or the velocity gradient.  
This is a curious fact,
clearly seen in the spectra of the Weaver and Williams (1972) survey, 
for example.  Since we know that the neutral hydrogen is spread fairly
smoothly from the inner Galaxy to well outside the solar circle,
we might expect the peak brightness temperature to correspond to low values
of the
velocity gradient, as it naturally would if most lines of sight
were optically thin.  On the other hand, we know that most of the
H{\tt I} is in the warm medium (roughly 75 \%, Kulkarni 1983), so the
peak brightness temperature does not simply correspond to the 
cloud temperature, $T_{cool}$, since the warm phase gas contributes
most of the column density at any velocity.  It is clear from
emission-absorption spectrum pairs at low latitudes, like those
in figures 2-8, that the peak
brightness in emission does not always correspond to the 
velocities of high optical
depths; on the contrary, deep absorption lines often 
correspond to dips in the emission brightness (i.e. H{\tt I} self-absorption).
It is a combination of the opacity, $< \kappa >$,
the cloud temperature, $T_{cool}$, and the warm phase density,
$n_w$, that determines the peak brightness temperature.

On a line of sight for which $\rho \gtrsim 1$, the 21-cm line is
%sep02 changed R to rho
fairly optically thick, and the typical line of sight distance, $\Delta s$,
needed to achieve average optical depth $\tau = 1$ over 
some velocity width $\Delta V$ is 

\begin{equation}
\Delta s \ = \ \frac{\Delta V}{< \kappa >}
\end{equation}

\noindent
since 

\begin{equation}
< \kappa > \ \ = \ \ \frac{EW}{\Delta s} \ \ = \ \ 
\frac{\int_{\Delta V}{\tau \ dv}}
{\Delta s}
\end{equation}

\noindent If we assume that the warm phase gas has mean density $n_w$,
then its column density over this line of sight interval is

\begin{equation}
N_w \ \ = \ \ n_w \ \Delta s \ \ = \ \ \frac{n_w \ \Delta V}{<\kappa>}
\end{equation}

\noindent
which, if there were no absorption, gives a velocity integral 
for the 21-cm line emission of 

\begin{equation}
\int{T_{w,u} \ dv} \ \ = \ \ \frac{N_w}{C_H} 
\end{equation}

\noindent
where the subscript $u$ indicates that this is the unabsorbed value,
not what we actually see.
Assuming once again that this warm phase emission is
spread smoothly in velocity
due to its broad natural line width ($\gtrsim 10$ km s$^{-1}$), we
can take
\begin{equation}
\int_{\Delta V}{T_{w,u} \ dv} \ \ \simeq \ \ T_{w,u} \ \Delta V
\end{equation}

\noindent
which gives

\begin{equation}
T_{w,u}  \ \ = \ \ \frac{N_w}{C_H \ \Delta V} \ = \ \frac{n_w}{C_H \ <\kappa>}.
\end{equation}

\noindent
As in equation 3, we can use $C_H = 5.9 \times 10^{-4}$ to get
$T_{w,u}$ in K for $n_w$ in cm$^{-3}$ and $<\kappa>$ in km s$^{-1}$
kpc$^{-1}$.  For a typical value of $n_w$=0.25 cm$^{-3}$ this gives
$T_{w,u}\ \simeq \ \frac{400 K}{<\kappa>}$, or about 80 K if 
$<\kappa> \simeq 5$ km s$^{-1}$ kpc$^{-1}$.  Note that we have
not had to assume a value of $\Delta V$ for this calculation,
it can be anything between the typical absorption linewidth of
$\lesssim$5 km s$^{-1}$ and the typical warm phase linewidth $\gtrsim$10
km s$^{-1}$.

The brightness temperature we expect from regions with $\rho \gtrsim 1$
%sep02 changed R to rho
is different from this, however, since $T_{w,u}$ is partially 
absorbed by the cool clouds before it gets to us, and the cool
clouds themselves contribute some emission, given by
$T_{cool} \ (1 \ - \ e^{-\tau})$.  If the warm phase gas is
well mixed around and among the cool clouds, then its attenuated brightness
temperature is given by  

\begin{equation}
T_{w} \ \ = \ \ T_{w,u} \ \frac{1\ - \ e^{-\tau}}{\tau} 
\end{equation}

\noindent
Adding on the emission from the cool cloud(s),
$T_{cool} \ (\ 1\ - \ e^{-\tau} )$,
gives the total brightness temperature seen in emission 

\begin{equation}
T_B \ \ = \ \ T_{cool} \ \left( 1 - e^{-\tau} \right) \ + \ 
\ T_{w,u} \ \frac{(1\ - \ e^{-\tau})}{\tau}.
\end{equation}

\noindent
For our assumed $\tau \simeq 1$ this gives $\sim$0.6 ($T_{cool}
+ T_{w,u}$), or about 0.6 (65K + 80K) = 87K.  

In the alternative geometry assumed in section 4, where the warm
phase gas is partly in front of the absorbing cloud and partly
behind, we get 

\begin{equation}
T_B \ \ = \ \ T_{cool} \ \left( 1 - e^{-\tau} \right) \ + \ 
\ T_{w,u} \ \left[ 1 - \epsilon \left( 1 - e^{-\tau} \right) \right]
\end{equation}

\noindent
or, in the notation of equations 10 - 14, 

\begin{equation}
T_B \ \ = \ \ T_{cool} \ x \ + 
\ T_{w,u} \ \left( \ 1 \ - \ \epsilon \ x  \right).
\end{equation}

\noindent
For $\epsilon$=0.5 and $x$=0.63 ($\tau$=1) this gives 102 K.  In both
cases the contributions from the warm and cool phases are about equal
at about 50 K each.  Going to higher optical depths does not change 
these numbers much.  For $\tau \rightarrow \infty$ ($x \rightarrow 1$)
equation 26 predicts $T_B \rightarrow 115$ K and in equation 24
the peak brightness actually decreases to approach $T_{cool}\simeq$ 65K.
Of course the actual value of the peak brightness temperature 
will fluctuate around these values due to the relatively wide
distribution of values of $T_{cool}$.  The highest values we
see presumably correspond to the warmer clouds ($T_{cool} \gtrsim 100$ K)
and regions of relatively low $<\kappa >$, i.e. long path lengths
between clouds.

In face-on disk galaxies the peak brightness temperature is much less 
dependent on the warm phase gas, since our lines of sight 
have path lengths of only a few hundred 
parsecs through the neutral hydrogen layer before hitting (or
missing) a cloud. 
This explains why the brightness temperatures
of the H{\tt I} emission from nearby galaxies show such strong variation
from place to place (Braun 1997).  This "high brightness network"
probably traces the structure of the cloud phase of the medium, 
which may include a significant fraction of the warm gas with it.
Payne, Salpeter and Terzian (1983), estimate that only about 30\%
of the total H{\tt I} emission is from the widespread, ``intercloud'' 
medium, the rest is associated with the clouds, of which about
half must be cool and the rest warm.  Braun (1997) sees many 
examples of brightness temperatures above 150 K; the SMC shows
a similarly high peak brightness temperature of 179 K (Stanimirovic
et al., 1999).  The implication of the high ends of the histograms
in figures 14 and 17 is that there must be some clouds in the Milky
%sep02 in figures 16 and 19 is that there must be some clouds in the Milky
Way with $T_{cool}$ as high as 150 K or more, although they are
rare compared with the 50 to 100 K clouds.  If we could get outside
the Milky Way and look back, we might see those clouds unobscured
by the cooler, more common clouds, so that seen from outside
the Milky Way also may show brightness temperatures of 150 K or more. 
Braun (1997) also sees a radial variation in the brightness temperatures
of the ``high brightness network'', with brightness temperatures 
increasing with radius.  This may be due to a radial variation of
$T_{cool}$, but it could also be explained by a radial variation in
$<\kappa>$ similar to what we find for the Milky Way in section 3;
if $<\kappa>$ decreases with $R_{gal}$ then $T_{w,u}$ increases
with $R_{gal}$ in equation 22, which makes the peak brightness
temperatures increase.

%sep02 added the paragraph below
The recognition that the peak brightness temperature of the 21-cm line
at low latitudes results from a trade-off between absorption in the cool
phase and emission from both warm and cool was first described by
Baker and Burton (1975) in the context of radiative transfer models
to fit the low latitude survey data then available.  The role of 
the velocity gradient in the peak brightness temperature problem
is discussed further by Burton (1993).  The only innovation in
the discussion above is to use the measured values of $< \kappa >$
together with the mean H{\tt I} density to get a rough quantitative estimate
for the value of the peak brightness temperature, rather than making 
a full cloudy medium model of the radiative transfer like that of
Liszt (1983).  The opacity function, $< \kappa (r) >$, incorporates
the optical depth predictions of these models without making 
specific assumptions about clouds sizes or geometries.

\subsection{H{\tt I} Self-Absorption}

Low latitude surveys of the Galactic 21-cm emission like the SGPS and
the CGPS have turned up many examples of H{\tt I} self-absorption (HISA, 
see Gibson et al., 2000, Minter et al. 2001, and Li and Goldsmith,
2002 for recent reviews).  Earlier surveys with large single dish
telescopes had shown that HISA is relatively common at low latitudes
(Baker and Burton, 1979).  It is often, but not always, associated
with molecular line emission (Burton et al. 1978, Knapp 1974).
%sep02 sentences added for more complete background

HISA is not true self-absorption, but absorption by cool foreground
H{\tt I} clouds of the background 21-cm line emission from warmer, more
distant gas.  Although temperature determinations for these HISA
clouds are imprecise, typically they must be cooler than 40K, and
sometimes as cold as 10 to 15K to fit the observations.  Such cold
H{\tt I} may be inside molecular clouds, but this is not always the case.
Although we do not yet have a complete catalog of such clouds,
there are some large examples (Knee and Brunt 2001) which suggest
that they may be common in the ISM.  In this analysis there are some
four clouds whose apparent temperatures are below 20 K, and several
more between 20 and 40 K, depending on $\epsilon$.  The most obvious
is in box 4 toward G326.65+0.59 (-24 $<$ v $<$ -17 km s$^{-1}$ on figure 13).
%sep02 The cool phase temperature is $T_{cool}=45$K, but
%sep02 cool phase temperature indicated on figure 17 is $T_{cool}=45$K, but
The effect of the cloud is to cause a deep dip in the emission 
spectrum at these velocities.

In principle, any cloud whose optical depth profile corresponds to
a dip in the emission spectrum, rather than a peak, qualifies as
HISA.  In the notation of equation 14, this criterion amounts to
%sep02 s \ \equiv \ c_2 \ - \ \left( c_0 \ + \ c_1 \right) \ < \ 0

\begin{equation}
s \ \equiv \ c_2 \ - \ \epsilon \ \left( c_0 \ + \ c_1 \right) \ < \ 0
\end{equation}

\noindent
so that the dependence of $T_B$ on $x$ has a negative coefficient.
For the $\epsilon$=0.5 case, the median value of $s$ is -2.5; fully
67\% of the lines have slope $s <$0.  Thus HISA is in fact more common
than the alternative, i.e. more than half of all H{\tt I} clouds, defined
as distinguishable features in the optical depth spectrum, correspond
to dips rather than peaks in the emission spectrum.  The HISA clouds
which can be distinguished on the H{\tt I} emission survey maps 
must be a relatively small subset of this population, those which
are either particularly close or particularly cold, so that they
stand out in their obscuring effect on the background emission.

%sep02 added sentences below.
At low latitudes and velocities corresponding to the inner galaxy,
the HISA effect is amplified for clouds on the near side of the
sub-central point due to unrelated emission from gas at the same
velocity on the far side.  Depending on the latitude, this should
typically increase the expected value of $\epsilon$ from 0.5 to 
as high as 0.75.  Using $\epsilon$=0.75 changes the median value
of the slope $s$ above only slightly, from -2.5 to -2.3, but still
more than half of the clouds (55\%) show negative values of $s$.

\section{Conclusions}

H{\tt I} spectra from low latitude surveys of the inner galaxy can
only be understood in terms of radiative transfer in a partially
absorbing medium.  The gas is far from optically thin, and it is
far from isothermal, so neither of these simplifications can be
used to understand the brightness temperatures we see in emission.
In this paper we study emission-absorption spectrum pairs to 
try to understand the relationship between the optical depth 
and the brightness temperature
through various clouds along several lines of sight.

This paper concentrates on a few spectra in a relatively small
area of the Galactic plane in the fourth quadrant.  We hope that
the techniques described here will be useful to analyze spectra
toward many more continuum sources throughout the SGPS, and 
perhaps in other similar surveys as well.  Since the sample is
small, the results described here must be considered as only
representative; more precise, quantitative measurements of the
properties of the cool neutral medium will come from larger
samples of spectra.

The three quantities that determine the characteristics of the
emission spectra that we see in low latitude surveys are the
warm phase H{\tt I} density, $n_w$, the temperature of the cool clouds,
$T_{cool}$, and the average density of the cool medium, which
sets the value of $<\kappa>$.  This paper discusses the last two,
the first has been determined from earlier surveys with lower
resolution (see reviews by Burton, 1988, Dickey and Lockman, 1990).
With these three quantities in hand, we can understand the
peak brightness of the H{\tt I} emission, the fraction of atomic gas in 
the warm and cool phases, and the significance and abundance of
HISA features.

The major findings here are: \begin{itemize}

\item {The mean opacity, $<\kappa>$, increases as we go inward
through the Galactic disk, from about 5 km s$^{-1}$ kpc$^{-1}$ 
at the solar circle to as much as 25 km s$^{-1}$ kpc$^{-1}$
in the molecular ring.  This result is in agreement with the
finding of Kolpak et al. (2002) for the first quadrant,
%sep02 but different from the earlier result of Garwood and Dickey (1988).}
but different from the earlier result of Garwood and Dickey (1989).}

\item {The temperature of the H{\tt I} in the cool clouds, $T_{cool}$,
is { \bf lower}
than what we would derive from a simple division of the emission by
the absorption, $T_{sp}$.  This agrees qualitatively 
(but not quantitatively) with 
results of fitting high and intermediate latitude emission-absorption
spectrum pairs by Heiles and Troland (2002).  The cloud 
temperatures derived here
are similar but slightly higher than temperatures obtained
for H{\tt I} clouds in the SMC (Dickey et al. 2000)
and in the LMC (Mebold et al. 1997) using a similar method.
Though the median cloud temperature is around 65K, the 
distribution of $T_{cool}$ has a tail reaching to very cold
values (well below 40 K).  }

\item{Putting together the values of $<\kappa>$ and $T_{cool}$ 
we can derive a rough prediction for the peak brightness
temperature of the Galactic H{\tt I} emission, that agrees with
the $\sim$125K that is seen, independent of the velocity
gradient, in single dish emission surveys.}

\item{The slopes of the fitted curves in the $T_{em}$, (1-e$^{- \tau}$)
plane suggest that HISA is a very widespread phenomenon.  The 
striking examples of HISA clouds that we notice in emission
surveys are only the particularly cold or fortuitously placed
cases.}\end{itemize}

Extending these techniques to the full survey area will 
indicate whether these conclusions apply to the inner 
galaxy as a whole, and also give a much better determination
of the distribution of cool cloud temperatures.

\acknowledgements

We are grateful to the ATNF staff for their hospitality and
assistance while the SGPS data were taken.  In particular we
thank Robin Wark, John Reynolds, Mark Wieringa, and Bob Sault
for their support and help.  We are grateful to Butler Burton,
Carl Heiles, Dain Kavars, Michael Kolpak, Harvey Liszt, F. J. Lockman,
Peter Martin, Simon Strasser, and to the anonymous referee for valuable 
criticism and suggestions.  We are grateful to Carl Heiles and Tom Troland
for giving us their Arecibo data prior to publication.
This research was supported in part by 
the National Science Foundation (grant AST 97-32695 to 
the University of Minnesota) and NASA (GSRP grant NGT 5-50250 to 
N.M-G.).  

\references

Allen, R.J., 2001, in {\bf Gas and Galaxy Evolution}, ASP Conference
Proceedings, Vol. 240, eds. J. E. Hibbard, M. Rupen, and J. H. van Gorkom
[San Francisco: Astronomical Society of the Pacific] p. 331.\\

Baker, P.L. and Burton, W.B. 1975, Ap. J., 198, 281.\\

Baker, P.L. and Burton, W.B., 1979, A. and A. Supp., 35, 129.\\

Braun, R. 1997, Ap. J. 484, 637.\\

Burton, W.B., 1988, in Galactic and Extragalactic Radio Astronomy, 2nd ed., eds. G.L. Verschuur and K. Kellerman, (New York :  Springer-Verlag) p. 295.\\

Burton, W.B., 1991, in {\bf The Galactic Interstellar Medium}, Saas-Fe 
Advanced Course 21, eds. P. Bartholdi and D. Pfenninger
[Berlin:  Springer-Verlag] section 2.3, pp. 20-22.

Burton, W.B., Liszt, H.S., and Baker, P.L., 1978, Ap. J. Lett. 219, 67.\\

Caswell, J.L., and Haynes, R.F., 1987, A. and A. 171, 261.\\

Clark, B.G., Radhakrishnan, V., and Wilson, R.W., 1962, /apj 135, 151.\\

Clark, B.G., 1965, Ap. J., 142, 1398.\\

Deshpande, A.A., 2000, MNRAS, 317, 199.\\

Dickey, J.M., Kulkarni, S.R., Heiles, C.E., and van Gorkom, J.H., 1983, Ap. J. Supp. 53, 591.\\

Dickey, J.M. and Lockman, F.J., 1990, Ann. Rev. Astron. Astrophys., 28, 215.\\

Dickey, J.M., Mebold, U., Stanimirovi\`c, S., and Staveley-Smith, L. 2000 \apj 536, 756\\

Dickey, J.M., McClure-Griffiths, N.M., Stanimirovic, S., Gaensler, B.M., Green, A.J., 2001, Ap.J. 561, 264.\\

Faison, M.D., Goss, W.M., 2001, A.J. 121, 2706.\\

Fich, M., Blitz, L., and Stark, A.A., 1989, \apj 342, 272.\\

Field, G.B., 1958, Proceedings of the IRE, 46, 240.\\

Field, G.B., Goldsmith, D.W., and Habing, H.J., 1969, Ap. J. Lett. 155, L149.\\

Gaensler, B.M., Dickel, J.R., and Green, A.J., 2000, Ap. J. 542, 380.\\

%Sep02 Garwood, R.W. and Dickey, J.M., 1988, Ap. J., 338, 841.\\
Garwood, R.W. and Dickey, J.M., 1989, Ap. J., 338, 841.\\

Gibson, S.J., Taylor, A.R., Higgs, L.A., and Dewdney, P.E., 2000 \apj 540, 851 \\

Goss, W.M., Radhakrishnan, V., Brooks, J.W., and Murray, J.D., 1972, Ap. J. Supp. 24, 123.\\

Hagen, J.P., Lilley, A.E., and McClain, E.F., 1955, Ap.J. 122, 361.\\

Heiles, C., 2000, Tetons 4: Galactic Structure, Stars, and the Interstellar Medium, ASP Conf. Ser. 231, ed C.E. Woodward, M.D. Bicay, and J.M. Shull (San Francisco: Astonomical Society of the Pacific), p. 294.\\

Heiles, C., 2001, \apj 551, 105\\

Heiles, C. and Troland, T.H., 2002, Ap. J. in press (astro-ph 0207104).\\

Hughes, M.P., Thompson, A.R., and Colvin, R.S., 1971, Ap. J. Supp.,  23, 323.\\

Kalberla, P.M.W., Schwarz, U.J., and Goss, W.M., 1985, Astron. Astrophys. 144, 27.\\

Knapp, G.R. 1974 \aj 79, 527 \\

Knee, L.B.G., and Brunt, C.M., 2001, Nature, 412, 308.\\

Kolpak, M. A., Jackson, J.M., Bania, T.M., and Dickey, J.M., 2002, Ap. J. in press.\\

Kuchar, T.A. and Bania, T.M., 1990, Ap. J. 352, 192.\\

Kulkarni, S.R., 1983, Ph.D. Thesis, U.C. Berkeley.\\

Kulkarni, S.R. and Heiles, C., 1988, in Galactic and Extragalactic Radio Astronomy, 2nd ed., eds. G.L. Verschuur and K. Kellerman, (New York : Springer-Verlag) p. 95.\\

Li, D. and Goldsmith, P.F., 2002, \apj submitted.\\

Liszt, H.S., 1983, \apj 275, 163.\\

Liszt, H., 2001, A\&A 371, 698.\\

Minter, A.H., Lockman, F.J., Langston, G.I., and Lockman, J.A., 2001, \apj 555, 868.\\

McClure-Griffiths, N.M., Dickey, J.M., Gaensler, B.M., Green, A.J., Haynes, R.F., Wieringa, M.H., 2000, A.J. 119, 2828.\\

McClure-Griffiths, N.M., Green, A.J., Dickey, J.M., Gaensler, B.M., Haynes, R.F., Wieringa, M.H., 2001, Ap.J. 551, 394.\\

Mebold, U., Winnberg, A., Kalberla, P.M.W., and Goss, W.M., 1982, A\&A, 115, 223.\\

Mebold, U., Dusterberg, C., Dickey, J.M., Staveley-Smith, L., Kalberla, P., Muller, H., and Osterberg, J., 1997, Ap. J. Lett, 490, 65.\\

Payne, H.E., Salpeter, E.E., and Terzian, Y., 1983, Ap. J. 272, 540.\\

Press, W.H., Teukolsky, S.A., Vetterling, W.T., and Flannery, B.P., 1992, ``Numerical Recipes in Fortran. The art of scientific computing'' (Cambridge:  University Press) 2nd ed., p.678.  \\

Radhakrishnan, V., Murray, J.D., Lockhart, P., and Whittle, R.P.J., 1972a, Ap. J. Suppl.  24, 15.\\

Radhakrishnan, V., Goss, W.M., Murray, J.D., and Brooks, J.W., 1972b, Ap. J. Supp. 24, 99.\\

Radhakrishnan, V., Murray, J.D., Lockhaart, P., and Whittle, R.P.J., 1972c, Ap. J. Supp. 24, 15.\\

Spitzer, L. Jr., 1977, {\bf Physical Processes in the Interstellar Medium}, (New York :  John Wiley), chap. 3.\\

Stanimirovic, S., Staveley-Smith, L., Dickey, J.M., Sault, R.J., and Snowden, S.L., 1999, MNRAS 302, 417.\\ 

Strasser, S. et al. 2002 in preparation.\\

Taylor, A.R., et al. 2002 in preparation.\\

Taylor, J.H. and Cordes, J.M., 1993, Ap. J. 411, 674.\\

Weaver, H. and Williams, D.R.W., 1973, Astron. Astrophys. Suppl., 8, 1.\\
  
\appendix
\section{Least Squares Solution}

The best fit values of $c_0$, $c_1$, and $c_2$ in equation (14) are
%sep02 fixed eq. no.
given by linear combinations of moments of the form shown in equation
(15).  The derivation of these formulae is very similar to the problem
%sep02 fixed eq. no.
of finding the coefficients of the best fit polynomial.
Begin by defining nine combinations of the moments :

\begin{equation}Y_1 \ = \ \overline{y} \ - \ \epsilon \ \overline{xy}
\end{equation}
\begin{equation}Y_2 \ = \ \overline{vy} \ - \ \epsilon \ \overline{vxy}
\end{equation}
\begin{equation}Y_3 \ = \ \overline{xy} 
\end{equation}
\begin{equation}K_1 \ = \  1 \ - \ 2 \epsilon \ \overline{x}
\ + \ \epsilon^2  \ \overline{x^2}
\end{equation}
\begin{equation}K_2 \ = \ \overline{v} \ - \ 2 \epsilon \ \overline{vx}
\ + \ \epsilon^2  \ \overline{v \ x^2}
\end{equation}
\begin{equation}K_3 \ = \ \overline{x} \ - \ \epsilon \ \overline{x^2}
\end{equation}
\begin{equation}K_4 \ = \ \overline{v^2} \ - \ 2 \epsilon \
\overline{v^2 \  x} \ + \ \epsilon^2  \ \overline{v^2 \ x^2} \end{equation}
\begin{equation}K_5 \ = \ \overline{vx} \ - \ \epsilon \ \overline{v \ x^2}
\end{equation}
\begin{equation}K_6 \ = \  \overline{x^2}
\end{equation}

then combine these to make four mixed quantities :

\begin{equation} H_1 \ = \ \left( Y_3 \ - \ \frac{Y_1 \ K_3}
{K_1} \right) \left( K_4 \ - \ \frac{K_2^2}{K_1} \right) \end{equation}

\begin{equation} H_2 \ = \ \left( Y_2 \ - \ \frac{Y_1 \ K_2}
{K_1} \right) \left( K_5 \ - \ \frac{K_2 \ K_3}{K_1} \right) \end{equation}
\begin{equation} H_3 \ = \ \left( K_6 \ - \ \frac{K_3^2}
{K_1} \right) \left( K_4 \ - \ \frac{K_2^2}{K_1} \right) \end{equation}

\begin{equation} H_4 \ = \ \left( K_5 \ - \ \frac{K_2 \ K_3}
{K_1} \right)^2 \end{equation}

and finally combining these we get simply 

\begin{equation}c_2 \ = \ \frac{H_1 \ - \ H_2}{H_3 \ - H_4}\end{equation}
\begin{equation}c_1 \ = \ \frac{Y_2 - \frac{K_2 \ Y_1}{K_1} \ - \ 
c_2 \ \left( K_5 \ - \ \frac{K_2 \ K_3}{K_1} \right)}
{K_4 \ - \ \frac{K_2^2}{K_1}}           \end{equation}

and

\begin{equation}c_0 \ = \ \frac{Y_1 \ - \ K_2 c_1 \ - \ K_3 c_2}
{K_1}\end{equation}

%\appendix
\section{Comparison with the Method of Heiles and Troland}

Using the Arecibo telescope, Heiles and Troland (2002a, b) have
recently carried out a monumental survey of emission and 
absorption in the 21-cm line toward some 75 bright continuum
sources at high and intermediate Galactic latitudes.  This
survey is complementary to the low latitude surveys because
it samples primarily gas within two kpc of the sun, but it
samples this nearby gas very well, since the continuum sources
are bright and even with the 3\arcmin \ (FWHM) beam of the
Arecibo dish the spatial resolution at the distance of the
gas is typically less than one parsec.  We have analysed their
spectra using the method described in section 4.3 above, to find
the distribution of cool phase temperatures shown on figure 18.
Counting each cloud (absorption feature) equally gives median
temperature 31 K for this distribution.  Weighting by the number
of velocity channels in each cloud raises this median temperature
to 35 K.

\begin{figure}
\vspace{2.5in}
\epsfig{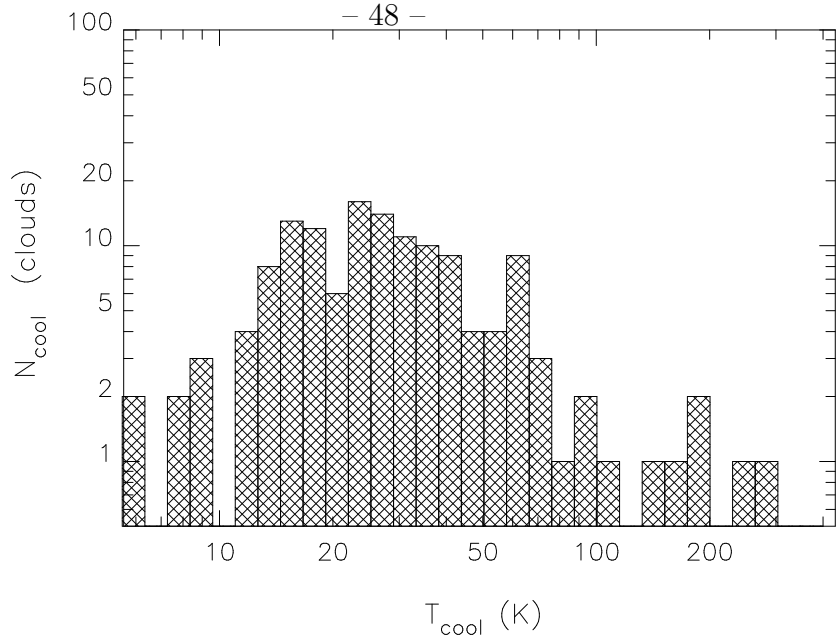}
\caption{Distribution of cool cloud temperatures, $T_{cool}$, 
for the absorption lines in the survey of Heiles and Troland
(2002a).  The median cool phase temperature is 31 K.}
\end{figure}

This result is somewhat cooler than the median temperature found
by Heiles and Troland (48K), so it is worth looking in detail at the
difference between their analysis and ours in a few 
pairs of spectra.  Table 4 from Heiles and Troland (2002a) gives
a comparison between their method and a simpler version of our
technique, which they call the ``slope method'' (Mebold et al. 1997).
For two sources they take $\epsilon \simeq 0.5$ on this table, 
3C333 and PKS0428+20 (which they correctly identify by this name on
their table 2).  For 3C333 they measure 29 K for
$T_{cool}$ from the slope method, and 25 K by fitting Gaussians,
whereas we measure 26 K using method 2 of section 4.3 above.  For
PKS0428+20 they measure 41 K and 2 K from the slope method for the
two dominant features centered at +3 and +10 km s$^{-1}$ respectively
(figures 19 and 20), while we find 38 K and 27 K. 
From Gaussian fitting, Heiles and Troland find 29 K for the first and
two blended components at 27 K and 34 K for the second.  These 
differences are minor, and they probably result from the apportionment
of the warm phase emission between the warm and cool components of 
Heiles and Troland's Gaussian analysis.

\begin{figure}
\vspace{2.9in}
\epsfig{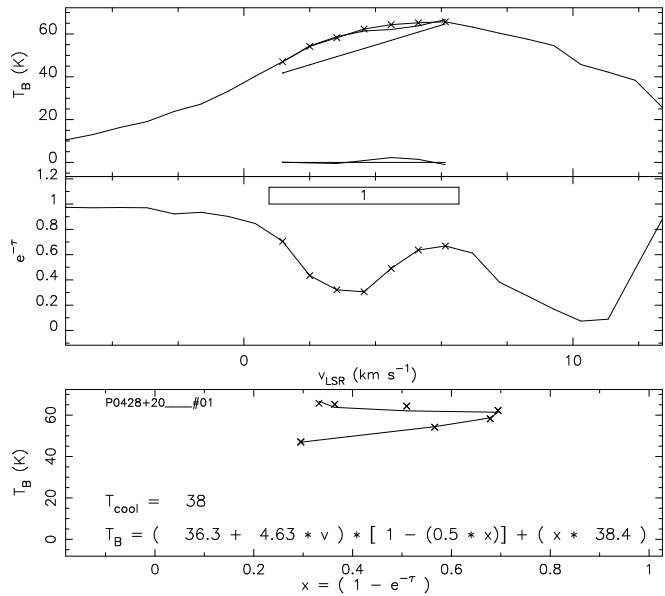}
\caption{Details of the fitting of the emission toward the
lower velocity absorption component of PKS0428+20.  Heiles
and Troland find 29 K for the cool phase temperature, while
we find 38 K.}
\end{figure}

\begin{figure}
\vspace{2in}
\epsfig{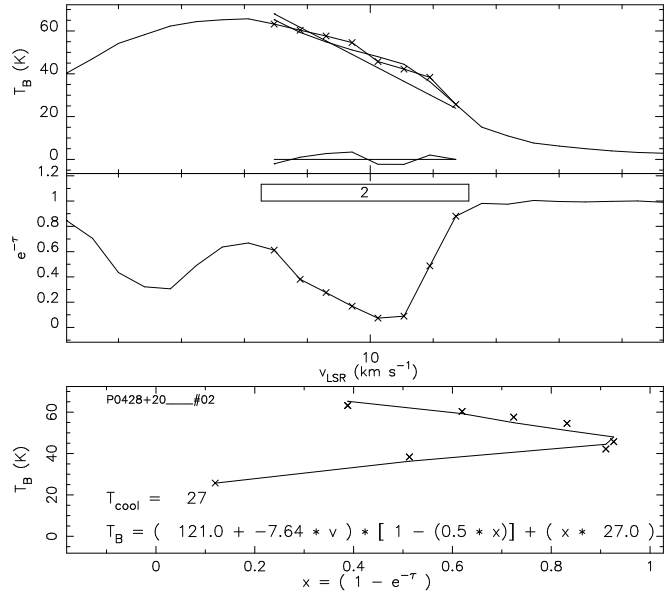}
\caption{Details of the fitting of the emission toward the
higher velocity absorption component of PKS0428+20.  Heiles
and Troland decompose this into two Gaussians with temperatures
27 K and 34 K, while we find 27 K for the combination.}
\end{figure}

The analysis method used by Heiles and Troland is to first decompose the
absorption spectrum into the sum of several Gaussians (three in the case
of PKS0428+20), and then to decompose the emission spectrum into the sum of
several more Gaussians (five in this case), of which the centers and
widths of the first three are set by the values obtained from fitting
the absorption.  This technique accomplishes the all-important 
separation of the emission into warm and cool components, where the
sum of the emission Gaussians that are {\bf not} seen in the absorption
correspond to $T_{w,f}\ + \ T_{w,b}$ in equation 13.  

There is no "right answer" to the problem of fitting multiple Gaussians,
the result depends on the number of components and the first guesses for
their parameter values.  The linear least squares method has this
advantage, that the solution for $T_{cool}$ and $T_w$ is unique 
given only an assumed value of $\epsilon$ and the velocity ranges
to use for the fitting.  At low latitudes, the approach of Heiles and 
Troland is not an option, since the emission spectra are so
complicated that Gaussian fitting is hopeless.  In the higher 
latitude directions studied in the Arecibo survey it is possible
to decompose the emission as well as the absorption into a few
Gaussians, and this gives more physical information about the gas
than the simple two-phase approximation used in this paper.  
%It is
%still the case, however, that the absorption spectra can be described
%by a smaller number of Gaussian components than are needed for the
%emission.  More important, 
In general the contribution to the emission by the 
components that are seen in absorption is less than
the contribution from those that are {\bf not} seen
in absorption, i.e. the WNM components; typically more than
half of the emission integral (column density) comes from Gaussians 
that are not visible in absorption, and so must be warmer than 150 to
200 K.  There is a good qualitative agreement on this result between
this paper and those of Heiles and Troland, in spite of some minor 
disagreements on specific values of $T_{cool}$ given by the two methods.

All methods of analysis seem to agree on the result that the typical
temperatures of absorbing clouds in the high and intermediate latitude
sample observed from Arecibo are cooler than the inner galaxy clouds
from the SGPS studied in this paper.  This may reflect a real gradient
in the cool phase temperature, either with Galactic radius, $R$, or with
height above the plane, $z$.  This may result from the different conditions
inside cloud complexes, i.e. the larger environments containing cool atomic
and molecular clouds and H{\tt II} regions.  The Arecibo sample contains
H{\tt I} clouds that are mostly not associated with H II region complexes,
they are in a typical interstellar radiation field.
If more of the H{\tt I} in the inner galaxy is recent photodissociation 
product due to the greater abundance of hot stars as suggested by
Allen (2001), then the typical cool phase temperature might be 
elevated, or there might be still more blending of warm, cool, and
intermediate temperatures that could bias our results to higher 
values of $T_{cool}$.  If further study of the SGPS data confirms
that there is a trend toward lower cool phase temperatures in
HI clouds with increasing $R$, then the corresponding {\bf decrease}
in $< \kappa >$ with $R$ suggests that the abundance of cool
clouds is decreasing rapidly with increasing $R$.

\end{document}